\documentclass[aps,prl,twocolumn,showpacs,superscriptaddress,10pt]{revtex4-1}

\usepackage{amsmath,amssymb,amsfonts}
\usepackage{graphicx}
\usepackage{color,xcolor}
\usepackage{mathtools,mathrsfs}
\usepackage[normalem]{ulem}
\usepackage{scalerel}

\usepackage{braket}
\usepackage{hyperref}

\begin{document}
\title{Finite-frequency prethermalization in periodically driven ergodic systems}
\author{Lorenzo Correale}
\affiliation{SISSA --- International School for Advanced Studies, via Bonomea 265, I-34136 Trieste, Italy}
\affiliation{INFN --- Istituto Nazionale di Fisica Nucleare, Sezione di Trieste, I-34136 Trieste, Italy}
\author{Leticia F. Cugliandolo}
\affiliation{Sorbonne Université, Laboratoire de Physique Théorique et Hautes Energies,
CNRS UMR 7589, 4 Place Jussieu, 75252 Paris Cedex 05, France}
\affiliation{Institut Universitaire de France, 1 rue Descartes, 75005 Paris France}
\author{Marco Schir{\`o}}
\affiliation{JEIP, UAR 3573 CNRS, Coll\`ege de France, PSL Research University, 11 Place Marcelin Berthelot, 75321 Paris Cedex 05, France}
\author{Alessandro Silva}
\affiliation{SISSA --- International School for Advanced Studies, via Bonomea 265, I-34136 Trieste, Italy}

\begin{abstract}
    We investigate the periodically driven dynamics of many-body systems, either classical or quantum, finite-dimensional or mean-field, displaying an unbounded phase-space.
    Using the lattice $\phi^4$ model and the $p$-spin spherical model as representative examples, we find that the inclusion of a smooth periodic drive atop an otherwise ergodic dynamics leads to a long-lived prethermalization, even at moderate driving frequencies.
    In specific asymptotic limits, we compute the corresponding prethermal Hamiltonian from an analytical perturbation scheme. 
\end{abstract}


\date{\today}
\maketitle

\emph{Introduction} ---
Periodically-driven systems 
are known to host a great deal of nontrivial phenomena. 
For example, an inverted pendulum can be stabilized by an external force oscillating at sufficiently high frequency~\cite{Kapitza1951}, while a kicked rotor exhibits an integrability-to-chaos transition~\cite{Chirikov1969stand_map} that is counter-intuitively suppressed by the inclusion of quantum fluctuations~\cite{Fishman1982localization}. 
Similarly, a driven quantum single particle system may localize in energy space resulting in bounded heating~\cite{Gefen}. 
While single-particle systems have been the focus of research in the past decades, attention has recently shifted towards many-particle ones, which have proven to be valuable tools to reproduce experimentally a wide array of novel phases of matter , within the field of \emph{Floquet engineering}~\cite{oka2019floquet}.
However, a major obstacle to their practical realization has been the fact that periodically driven isolated many-body systems tend to quickly heat up to a featureless infinite-temperature state, due to the absence of conservation laws~\cite{DAlessio2014longtime,Lazarides2014equilibrium,Ponte2015mbl,Genske2015floquetboltzmann}. 

Infinite-temperature thermalization can be avoided by many-body localization ~\cite{Khemani2021driven,Else201floquet,Yao2017dtc} and integrability~\cite{Russomanno2012ising,Citro2015kapitza,Gritsev2017integrable,Ishii2018integrable,Prosen1998time_evol,Prosen1998quantum_invariants,Prosen1999ergodic_properties}.  
For generic short-range interacting systems, heating can be suppressed only in a finite time window.
For example, a prethermalization induced by a high-frequency drive $\Omega$ has been observed in numerical studies both in classical~\cite{Howell2019,Pizzi2021prethermal,Sadia2022chaos_rotors} and quantum systems~\cite{Kuwahara2016magnus,Mori2016bound,Abanin2017rigorous_preth,Abanin2017effective_hamiltonians,peronaciPRL2018,PeronaciEtAlPRB2020,
Ho2023prethermalization,dallatorre2021statistical}, as well as in some experimental setups~\cite{neyenhuis2017observation,Rubio2020bosehubbardfloquet,peng2021floquet,shkedrov2022absence}. In this case, the corresponding prethermal Hamiltonian is asymptotically close to the average over a period, $\overline{H(t)}$, of the true periodic Hamiltonian $H(t)$. While most literature has concentrated on systems with a locally bounded energy spectrum, high-frequency prethermalization has been conjectured as a generic outcome also for unbounded spectra~\cite{Hodson2021diffusion}, as supported by numerical studies on periodically kicked systems~\cite{Rajak2018preth_rotors,Rajak2019,Sadia2022chaos_rotors} and experiments involving a uniform Fermi gas~\cite{shkedrov2022absence}. 

From a broader perspective, prethermalization is also observed close to an integrable point, both in quenched systems~\cite{Mori_2018} as well as under periodic drive ~\cite{Knapp2017prethermal_keldysh}. 
Beyond these scenarios, the conventional expectation is that a non-integrable system with intermediate interaction couplings and drive frequency should rapidly thermalize. In this work, we show that this expectation is not met when a system with an unbounded energy spectrum is driven under a smooth periodic force.

{We investigate two systems, namely the 1D lattice $\phi^4$ model and the $p$-spin spherical model (PSM). For the parameters we choose, the dynamics of both models is ergodic and rapidly relaxes to thermal equilibrium in absence of the drive. 
Instead, when coupled to a finite-frequency drive, both systems display a long-lived prethermalization that can not be traced back to an high-frequency expansion. 
For the $\phi^4$ model, we calculate the corresponding prethermal Hamiltonian ${\mathcal H}_F$ in an asymptotic, though not high-frequency, limit. This calculation reveals that prethermalization results from a trade-off between resonant energy absorption and non-linearity.}
{Additionally, we show that both classical and quantum PSMs approximately obey a classical fluctuation-dissipation relation, supporting the likelihood of approaching a prethermal finite-temperature state.}


\begin{figure*}[t]
    \centering
    \includegraphics[width=0.85\textwidth]{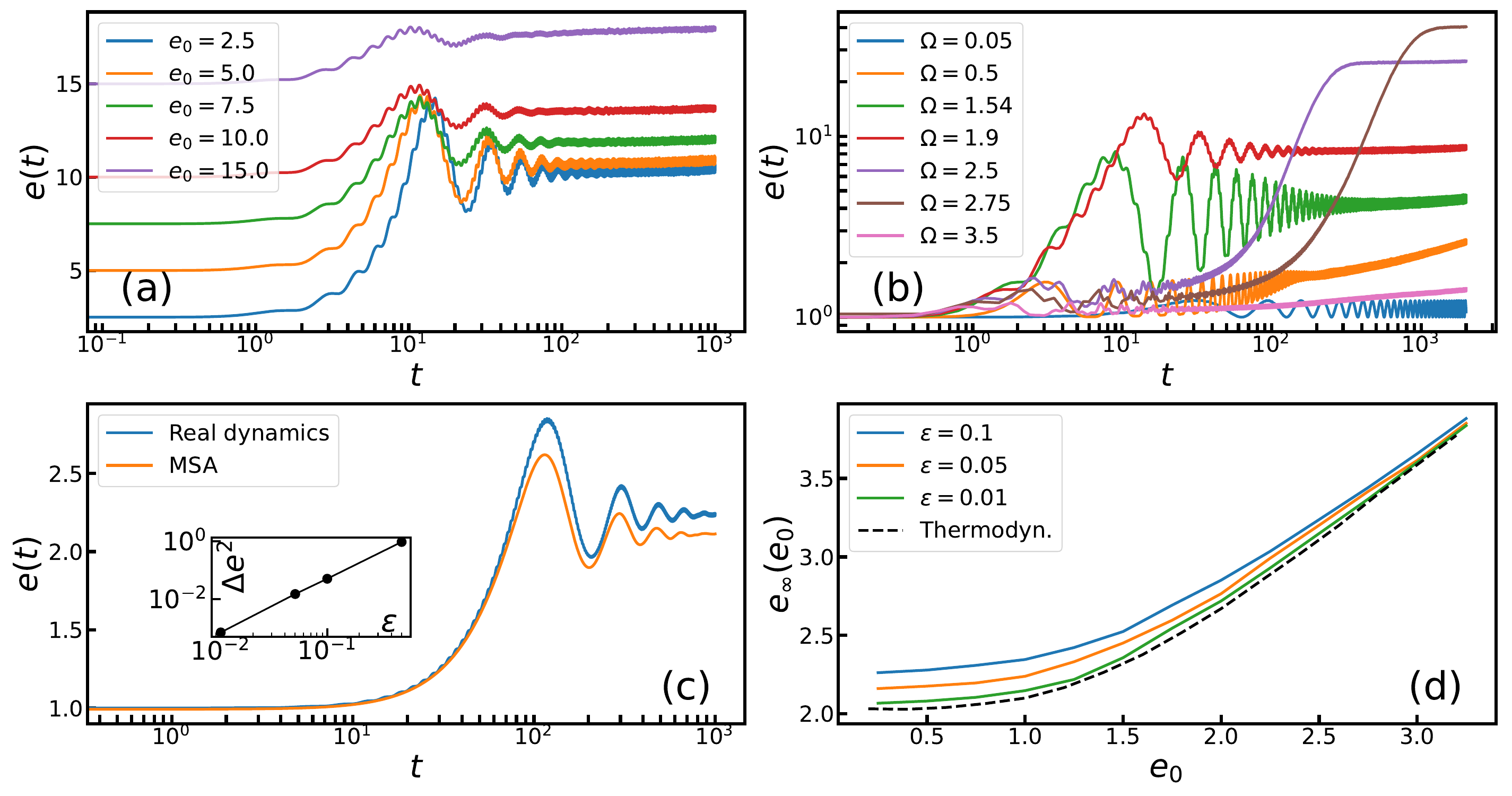}
    \caption{(Color online) One dimensional lattice $\phi^4$ model with $N=100$. Data averaged over $\mathcal{N}=1000$ configurations. 
    (a)-(b) The energy density $e(t)=\braket{H_{\phi^4}\big(\mathbf{x}(0),\dot{\mathbf{x}}(0)\big)}_{H_0}/N$, obtained by integrating Eqs.~\eqref{eq:phi4_Hamilton_eq}, with $B_0=g=\lambda=\omega_0^2=1$. 
    %
    %
    (a) $\Omega=2.4$ and several $e_0$.
    (b) $e_0=1$ and several $\Omega$. {In particular, in the red and the green plot, $\Omega$ is fixed to a normal mode frequency $\omega_k$, respectively for $k=4\pi/5$ and $k=3\pi/5$.}
    %
    (c) Comparison between $e(t)$ and the energy density in Eq.~(\ref{eq:phi4_estimate_energy_density}) estimated from the effective dynamics in Eqs.~\eqref{eq:average_eq_u&v_phi4}, 
    with $e_0=\Omega=\tilde{B}_0=\tilde{g}=\tilde{\lambda}=1$, $\Delta\omega^2=3$, $\epsilon=0.05$. The parameters of Eq.~\eqref{eq:phi4_Hamilton_eq} are determined by the rescaling in the main text. Inset: time-averaged difference $\Delta e^2$, between the real and the MSA dynamics, as defined in the main text. To avoid transient effects, we perform the time-average over a window $[3/4~t_{max},t_{max}]$, with $t_{max}$ the maximum simulation time.
    (d) $e_0$ dependence of 
    the asymptotic energy density $e_\infty$, computed as the average Eq.~\eqref{eq:E_inf_time_avg}
    over the same late time window of the dynamics in Eqs.~\eqref{eq:phi4_Hamilton_eq}. The dependence on $\epsilon$ is introduced through $\tilde B_0, \tilde g, \tilde \lambda$.
    The dashed line corresponds to the microcanonical estimation $e_\infty^{\rm MSA}(e_0)$, obtained from the MSA as described in the main text.}
    \label{fig:letter_phi4}
\end{figure*}

\emph{The classical 1D lattice $\phi^4$ model}  --- The model is
\begin{equation}
\label{eq:phi4_Hamiltonian}
     H_{\phi^4}(\mathbf{x},\mathbf{p})  =  \sum_{i=1}^N \bigg[\frac{p_i^2}2 +  \frac {\omega_0^2}2 x_i^2 + \frac {\lambda}4 x_i^4  + \frac g2 (x_{i+1}-x_i)^2\bigg] 
\end{equation}
with $\omega_0^2>0$ and periodic boundary conditions. {The normal modes of its quadratic part have frequencies $\omega_k = \sqrt{\omega_0^2 + 2g[1-\cos(2\pi k/N)]}$, for $k=0\ldots N-1$}.
In absence of an external drive and for the parameters we consider here, the Hamiltonian dynamics displays ergodicity~\cite{Parisi1997phi4_ergodic,Aarts2000phi4_ergodic} (see also the SM~\cite{SM}) and chaos~\cite{Pettini1991sst,Pettini1993geometricalhints}.
Several experimental realizations have been proposed for its quantum extension~\cite{Shimshoni2011zigzag,Silvi2013zigzag,Silvi2014dm_zigzag}.
We aim to study the dynamics driven by the term 
$H_1(\textbf{x},t)=-B_0 \sin(\Omega t)\sum_i x_i$, {summarized in the classical equations of motion:}
\begin{equation}\label{eq:phi4_Hamilton_eq}
    \ddot{x}_i + \omega_0^2 x_i = B_0 \sin( \Omega t) + g(x_{i+1} -2 x_i + x_{i-1}) -\lambda x_i^3 \ ,
\end{equation}
for $i=1,\dots,N$.
In Fig.~\ref{fig:letter_phi4} we plot the energy density, $e(t)=\braket{H_{\phi^4}(\textbf{x}(t),\dot{\textbf{x}}(t))}_0/N$, averaged over $\mathcal{N}$ initial conditions randomly sampled on $H_{\phi^4}\big(\mathbf{x}(0),\dot{\mathbf{x}}(0)\big) = Ne_0$, a 
constant energy manifold.
For all the explored values of $e_0$ (a) and for $\Omega$ (b), $e(t)$ saturates to a finite $e_\infty$ at long times, signalling prethermalization.
{The plateau at $e_\infty$ is quickly approached when $\Omega$ matches or is close to one of the normal frequencies $\omega_k$, suggesting that this finite-frequency prethermalization arises from the interplay between a resonant drive and the non-linear force $-\lambda x_i^3$. An analogous mechanism leads to an effective conservation law in the single-body version of this model, namely the Duffing oscillator~\cite{scholarpedia_duffing_oscillator}. Conversely, when $\Omega$ is outside the range of normal frequencies, $e(t)$ exhibits another early-time plateau, located close to $e_0$, likely due to adiabatic evolution for low $\Omega$ and high-frequency prethermalization for high $\Omega$. This observation further confirms that the plateau at $e_\infty$ should not be attributed to high-frequency prethermalization. In the Supplemental Material (SM), we show that the finite-frequency plateau at $e_\infty$ has a finite life-time, controlled by $\lambda$~\cite{SM}. We also demonstrate that our results remain robust with different forms of $H_1(\textbf{x},t)$. However, the lifetime of the prethermal plateau decreases as $H_1(\textbf{x},t)$ becomes a less smooth, consistently with the expectation that finite-frequency prethermalization disappears in the limit where the drive is a periodic sequence of kicks~\cite{Rajak2018preth_rotors,Rajak2019,Sadia2022chaos_rotors}.}

{Since the origin of the prethermalization at $e_\infty$ can not attributed to high-frequency effects, we need an analytical approach different from the high-frequency (Magnus) expansion~\cite{Magnus1954expansion} to study this phenomenon. For this purpose, we employ a multiple-scale analysis (MSA), which has been successfully used to study single-body driven systems~\cite{Bender1999mathmethods,Fishman2003multiscale}. In the following, we apply MSA to the lattice $\phi^4$ model under study, although our analysis can be generalized to a wider class of many-body models (see also SM~\cite{SM}). We begin by defining an asymptotic limit where the prethermal time-scale is much larger than $\Omega^{-1}$. This is achieved through the rescaling $B_0=\epsilon\Tilde{B}_0$, $\lambda=\epsilon\Tilde{\lambda}$ and $g= \epsilon\Tilde{g}$, with $\epsilon \ll 1$. We also set $\omega_0=\sqrt{\Omega^2-\epsilon\Delta\omega^2}$, with $\Omega\sim O(1)$, to enable a parametric resonance even for small $\epsilon$.}
{Then, we introduce a redundant time variable $\tau=\epsilon t$, intuitively representing the `slow' prethermalization time-scale.}
We propose the asymptotic expansion
\begin{equation}\label{eq:multiscale_expansion}
    x_i(t) = x_{0,i}(t,\tau) + \epsilon \ x_{1,i}(t,\tau) + \dots 
\end{equation}
and treat $t$ and $\tau$ as they were \emph{independent}. Although the exact solution $\mathbf{x}(t)$ depends only on $t$, this trick is useful to eliminate secular terms at each perturbative order, {as shown in the following.} We plug Eq.~\eqref{eq:multiscale_expansion}, together with the chain rule $d/dt = \partial_t + \epsilon\partial_\tau$, into Eq.~\eqref{eq:phi4_Hamilton_eq}. After collecting powers of $\epsilon$, the first two perturbative orders read~\cite{SM}:
\begin{align}
    \left(\partial_t^2 + \Omega^2\right) x_{0,i} & \ =  \ 0 \ , 
    \label{eq:phi4_multiscale_0}\\
    \left(\partial_t^2 + \Omega^2\right) x_{1,i} &= -2 \partial_t\partial_\tau x_{0,i} + \Delta\omega^2 x_{0,i} ~ + \label{eq:phi4_multiscale_1} \\
    & \ \ \ - \Tilde{\lambda} x_{0,i}^3 + \Tilde{B}_0 \sin(\Omega t) 
    ~ . 
    \nonumber 
\end{align}
The leading order solution of Eq.~\eqref{eq:phi4_multiscale_0} is given by $ x_{0,i}(t,\tau) = u_i(\tau)\cos(\Omega t) + v_i(\tau) \sin(\Omega t)$. The slowly varying amplitudes are determined by imposing that secular terms, proportional to $\sin(\Omega t)$ or $\cos(\Omega t)$, are absent in the right-hand-side of Eq.~\eqref{eq:phi4_multiscale_1}, the next order, as these {would rule out the presence of a plateau in the dynamics, contradicting our numerical results}. After some algebra~\cite{SM}, one finds the Hamiltonian equations
\begin{equation}\label{eq:average_eq_u&v_phi4}
    \frac{du_i}{d\tau} = -\frac{\partial \mathcal{H}_F}{\partial v_i}\ , \qquad\;\;\;\;
    \frac{dv_i}{d\tau} = \frac{\partial \mathcal{H}_F}{\partial u_i}
    \; , 
\end{equation}
generated by the many-body, local effective Hamiltonian
\begin{eqnarray}
    &\mathcal{H}_F(\mathbf{u},\mathbf{v})= \frac 1 {2\Omega}\sum_i \Big\{ \! -\frac {\Delta\omega^2}2 (u_i^2 + v_i^2)  +
    \frac {\tilde{g}}2 \Big[(u_{i+1}-u_i)^2 
    \nonumber\\
    &+(v_{i+1}-v_i)^2  \Big] 
    + \frac {3 \tilde{\lambda} } {16} (u_i^2+v_i^2)^2 + \Tilde{B}_0 v_i  \Big\} 
    \ . \label{eq:phi4_eff_hamiltonian}
\end{eqnarray}
{Once the dynamics of $\mathbf{u}(\tau)$ and $\mathbf{v}(\tau)$ is solved, the MSA allows us to predict quantitatively the time evolution of the energy density.} By plugging the leading order solution $x_{0,i}(t,\tau)$ into Eq.~\eqref{eq:phi4_Hamiltonian}, we estimate $e(t)$ as
\begin{align}
    e^{\rm MSA}(t)
    & 
    =
    \frac 1N\braket{H_{\phi^4}[\mathbf{x}_0(t,\tau), \partial_t\mathbf{x}_0(t,\tau)]}_0 \Big|_{\tau=\epsilon t} = \nonumber \\
    &=\frac 1 N \braket{\mathcal{H}_0[\mathbf{u}(\tau),\mathbf{v}(\tau)]}_0\Big|_{\tau=\epsilon t} + O(\epsilon) \label{eq:phi4_estimate_energy_density}
    \ .
\end{align}
Here, $\mathcal{H}_0(\mathbf{u},\mathbf{v})=\Omega^2\sum_i (u_i^2+v_i^2)/2$ is obtained omitting the order $\epsilon$ terms in $g$ and $\lambda$ from Eq.~\eqref{eq:phi4_Hamiltonian}.
The average of $\braket{\cdot}_0$ is still performed over microcanonically sampled initial conditions, given by $\mathbf{x}_0(0)=\mathbf{u}(0)$ and $\dot{\mathbf{x}}_0(0)=\mathbf{v}(0)/\Omega$.
In Fig.~\ref{fig:letter_phi4}(c), we compare the estimate from Eq.~\eqref{eq:phi4_estimate_energy_density} with the results from the full dynamics in Eqs.~\eqref{eq:phi4_Hamilton_eq}.
The MSA successfully reproduces the dynamics up to the {prethermal time-scale}. {Furthermore, as shown in the inset, the time averaged error $\Delta e^2 =  \int_{t_0}^{t_0+\mathcal{T}} dt |e(t)-e^{\rm MSA}(t)|^2/ \mathcal{T}$ consistently vanishes for vanishing $\epsilon$, as long as the maximum integration time $\mathcal{T}$ is smaller than the thermalization time.}

{The main conclusion from MSA is that the long-time plateau in $e^{\rm MSA}(t)$ can be analytically understood as a prethermalization in a microcanonical ensemble, determined by the effective Hamiltonian from Eq.~\eqref{eq:phi4_eff_hamiltonian}. As we anticipated, the effective Hamiltonian $\mathcal{H}_F(\mathbf{u},\mathbf{v})$ results from the interplay between a resonant field  and a non-linearity, represented in Eq.~\eqref{eq:phi4_eff_hamiltonian} by the terms proportional to $\Tilde{B}_0$ and $\Tilde{\lambda}$, respectively. We observe that, differently from single-body systems, here the MSA captures the correct behaviour of the dynamics only up to a finite time,  as it does not account for the further heating that follows prethermalization. This discrepancy may be either due to the leading order truncation in the calculation or to a vanishing radius of convergence of the MSA in the thermodynamic limit, as observed also for the high-frequency expansion~\cite{DAlessio2014longtime}. Further investigation into this discrepancy is left for future work.}

{Based on the previous observations, from MSA we can also quantitatively predict the value of the prethermal plateau $e_\infty$, as a function of the initial energy density $e_0$, using equilibrium statistical mechanics only. Specifically, we first observe that $e^{\rm MSA}(t)$ is expected to relax to the equilibrium average $e^{\rm MSA}_\infty(e_F)=\braket{\mathcal{H}_0(\mathbf{u},\mathbf{v})}_F/N$, over the prethermal microcanonical ensemble determined by $\mathcal{H}_F(\mathbf{u},\mathbf{v}) =N e_F$. The effective energy density $e_F$ is in turn determined by the initial condition of dynamics through the average $e_F(e_0)=\braket{\mathcal{H}_F(\mathbf{u},\mathbf{v})}_0/N$,
computed over a uniform distribution on the microcanonical shell $N e_0 = \mathcal{H}_0(\mathbf{u},\mathbf{v})$, which approximates the original initial conditions of our protocol, at leading order in $\epsilon$.
Combining $e^{\rm MSA}_\infty(e_F)$ and $e_F(e_0)$, we obtain the relationship $e^{\rm MSA}_\infty(e_0)$. 
We compare this result to the corresponding numerical estimation of of $e_\infty(e_0)$,} obtained from the original dynamics in Eq.~\eqref{eq:phi4_Hamilton_eq}, averaged over a late time window
\begin{equation}\label{eq:E_inf_time_avg}
    e_{\infty}(e_0) = \frac {1}{N\mathcal{T}} \int_{t_0}^{t_0+\mathcal{T}} dt  \braket{H_{\phi^4}(\mathbf{x}(t),\mathbf{p}(t))}_0 \ ,
\end{equation}
for the same initial conditions on the manifold $H_{\phi^4}(\mathbf{x},\mathbf{p})=N e_0$ {and for $t_0$ lying in the prethermalization time-window.}
The comparison, illustrated in Fig.~\ref{fig:letter_phi4}(d), demonstrates that the numerical estimation from Eq.~\eqref{eq:E_inf_time_avg} indeed converges to the equilibrium estimation $e^{\rm MSA}_\infty(e_0)$, for $\epsilon\to 0$.
Our analysis may be improved by including higher-order terms from Eq.~\eqref{eq:multiscale_expansion}.


\begin{figure*}[t]
\centering
\includegraphics[width=0.85\textwidth]{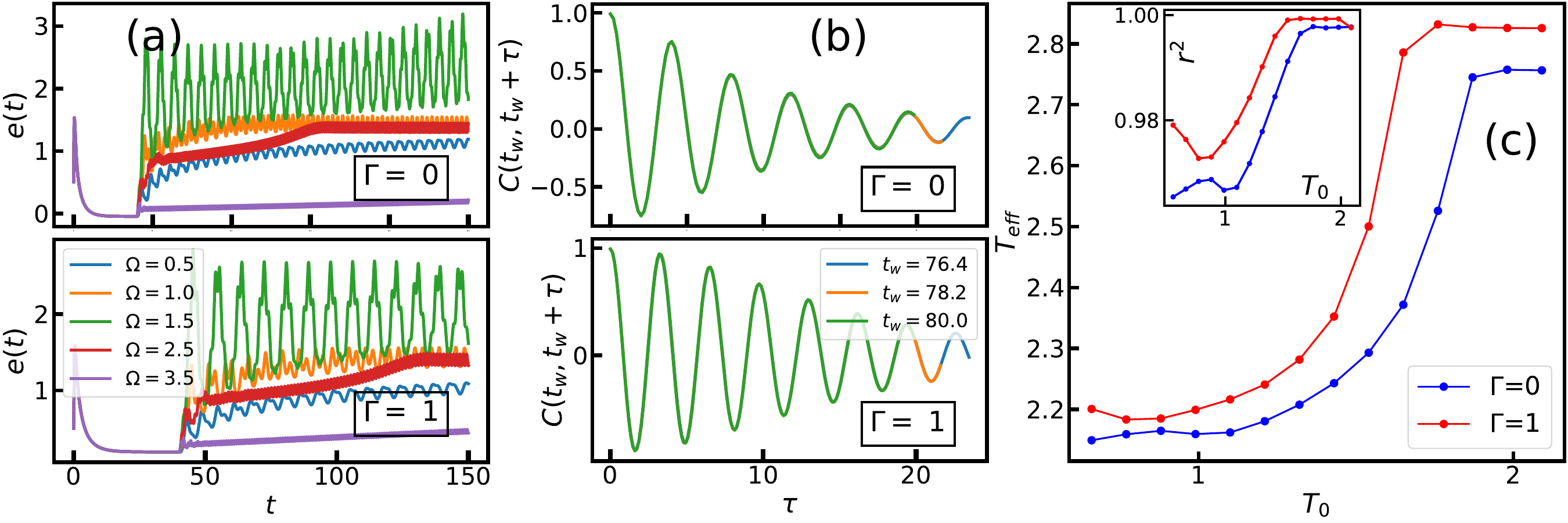}
    \caption{The classical ($\Gamma=0$) and quantum ($\Gamma=1$) PSM. $B_0=M=J=1$, $p=3$, $t_b=40$ and $T_0=1$. {With these parameters, we have $T_d(\Gamma=0)\simeq 0.6$ and $T_d(\Gamma=1)\simeq 0.5$~\cite{Cugliandolo2000firstorder}.}
    (a) The energy density $e(t)$, for several  values of $\Omega$. 
    (b) The symmetric correlation, Eq.~\eqref{eq:corr_def}, evolving from an initial state at temperature $T_0=1.98$. 
    (c) The effective temperature $T_{\text{eff}}$ against  $T_0$. The data are obtained through a linear regression of the averaged two-point functions $d\overline{C}(\tau)/d\tau$ and $\overline{R}(\tau)$ from Eqs.~\eqref{eq:letter_avg_corr} and \eqref{eq:letter_avg_resp}, using the classical FDT. The corresponding $r^2$ parameter is reported in the inset.}
    \label{fig:letter_PSM_results}
\end{figure*}

\emph{The classical and quantum fully-connected PSM}  --- We can now 
generalize our findings to classical and quantum systems of mean-field kind, 
focusing on the driven $p$-spin spherical model (PSM):
\begin{eqnarray}
\label{eq:hamiltonian_PSM}
\hat{H}_J 
\! = \!
    \sum_i \frac{\hat{\Pi}_i^2}{2M}  - 
    \!\!\!\! \!\! 
    \sum_{i_1 < ... < i_p} 
    \!\!\!\! \!\! 
    J_{i_1...i_p} \!\hat{\sigma}_{i_1}...\hat{\sigma}_{i_p}
\! -  \! B_0\sin(\Omega t)\sum_i\hat{\sigma}_i  \;
\end{eqnarray}
The PSM describes a gas of $N$ particles interacting through random, all-to-all couplings $J_{i_1,\dots,i_p}$, independently sampled from a Gaussian distribution with zero mean and variance $\overline{J^2}=p! J^2/2 N^{p-1}$. The particles are globally constrained over an $N$-sphere, $\sum_i \braket{\hat{\sigma}_i^2}=N$~\cite{kosterlitz1976spherical}. Quantum fluctuations are implemented by the canonical quantization relations $[\hat{\sigma}_i, \hat{\Pi}_j] = i \hbar \delta_{ij}$ 
and are controlled by the dimensionless parameter $\Gamma=\hbar^2/MJ$~\cite{Cugliandolo2000firstorder}. In the following, we will set $M=J=1$.
On a critical line $T_c(\Gamma)$, the PSM displays a thermodynamic phase transition between a paramagnetic and a glassy state, which is either of the first or second order depending on the strength of $\Gamma$~\cite{Cugliandolo2000firstorder,cugliandolo2001imaginary}.
From a dynamical point of view, the model is ergodic at high-temperatures and non-ergodic below a temperature $T_d(\Gamma)>T_c(\Gamma)$ ~\cite{cugliandolo1999real}. Evidences for dynamical and quantum chaos are retrieved in the ergodic and non-ergodic phases~\cite{bera2021quantum,anous2021quantum,Winer2022,Correale2023probing}.

We work at $T>T_d(\Gamma)$, where the PSM is  ergodic. The symmetric correlation and linear response function
\begin{align}
C(t,t') &= \frac12 \langle \hat{\sigma}(t)\hat{\sigma}(t')+\hat{\sigma}(t')\hat{\sigma}(t)\rangle \label{eq:corr_def} \; , 
\\
R(t,t') &= \frac{i}{\hbar}\theta(t-t')\langle[\hat{\sigma}(t),\hat{\sigma}(t')]\rangle \ , \label{eq:resp_def}
\end{align}
obey a closed set of Schwinger-Dyson equations, valid for $N\to\infty$ and under disorder average~\cite{cugliandolo1999real,thomson2020quantum,Berthier2001,Busiello2004drivenPSM}
(see the SM~\cite{SM}).
We initialize the dynamics
by evolving the undriven  ($B_0=0$) system in contact with a bath of harmonic oscillators at temperature $T_0>T_d$ during a finite time $0<t<t_b$~\cite{thomson2020quantum}. At $t_b$ we switch
off the coupling to the bath and we follow the periodically driven dynamics ($B_0>0$) of the isolated system. 

In Fig.~\ref{fig:letter_PSM_results}(a)
we plot  the average energy density $e(t)$~\footnote{$\lim_{N\to\infty} e(t)$ can be expressed in terms of $C(t,t')$ and $R(t,t')$ (see SM~\cite{SM}).} {for a fixed $T_0>T_d(\Gamma)$ and several values of $\Omega$.}
Initially, $e(t)$ saturates to a first plateau for $t\lesssim t_b$, signalling thermal equilibrium with the bath, and later oscillates around a second one at $e_\infty$.
{This second saturation suggests both classical and quantum PSM prethermalize.}
Like in the lattice $\phi^4$ model, the prethermalization is robust against variations in $\Omega$ and in the form of the driving force~\cite{SM}. {Within our simulation time-scale, we do not observe deviations of $e(t)$ from $e_\infty$ from the Schwinger-Dyson equations. However, in the SM, we show that the system eventually thermalizes to infinite temperature at least when $\Gamma=0$ and its size $N$ is finite, by integrating the Hamilton equations of motion for the PSM~\cite{SM}}.

{The prethermalization we observe in the PSM is reminiscent of that discussed for the lattice $\phi^4$ model. In both cases, prethermalization results from the interplay between resonance and non-linearity. For $J=0$, a characteristic frequency for the PSM can be identified with the inverse of the period of the rotations around the $N$-sphere. This feature is more transparent in the single-body version of the driven PSM, analyzed in the SM, where the effective conserved quantity is obtained through a MSA~\cite{SM}.
In the many-body scenario with $J>0$, this typical frequency broadens into a finite spectrum, identified with
the support of the equilibrium response function $\chi^{\prime\prime}(\omega)$, that is the transform of $R(t,t')$ of the undriven PSM~\cite{cugliandolo1999real}. However, unlike the $\phi^4$ model, we do not apply the MSA to the many-body driven PSM, as this would require using $N$-dimensional spherical coordinates, significantly complicating the analysis.}
Instead, as we have direct access to the two-point functions, we search for other indications of relaxation to a Gibbs-like state from the Fluctuation-Dissipation Theorem (FDT)~\cite{Nielsen1996fdt}. Its classical ($\Gamma=0$) version states that whenever a system is in equilibrium at temperature $T$,  
$dC(\tau)/d\tau = -T ~ R(\tau)$, with $\tau=t-t'>0$.
In Fig.~\ref{fig:letter_PSM_results}(b) we show that, although the dynamics is non-stationary due to the driving, $C$ and $R$ exhibit an \emph{approximate} discrete stationarity in the late-time regime, corresponding to the invariance 
under simultaneous translations of $t$ and $t'$ by the period $2\pi/\Omega$.
Motivated by this observation, we test the FDT for the averaged two-point functions:
\begin{eqnarray}
&& \!\!\!\!\!\!\!\!\!\!\!\!\!\!
    \overline{C}(\tau)=\frac{\Omega}{4\pi}\int_{t_0}^{t_0+\tfrac{2\pi}{\Omega}} 
    \!\! \!\! ds \left[C(s+\tau,s) - m(s+\tau)m(s)\right] 
     \label{eq:letter_avg_corr} 
\\
&& \!\!\!\!\!\!\!\!\!\!\!\!\!\!
    \overline{R}(\tau)=\frac{\Omega}{4\pi}\int_{t_0}^{t_0+\tfrac{2\pi}{\Omega}} \!\! \!\! ds \; R(s+\tau,s) \label{eq:letter_avg_resp} 
    \; , 
\end{eqnarray}
with $m(t)= \int_0^t ds \; R(t,s)B(s)$ the oscillating magnetization~\cite{SM}.
From the slope of a linear regression of $\overline{R}(\tau)$ against $d\overline{C}(\tau)/d\tau$, we obtain an \emph{effective temperature} $T_{\text{eff}}$ at late times.
The $T_0$ dependence of $T_{\text{eff}}$ is displayed, along with the corresponding $r^2$ parameter from the regression, in Fig.~\ref{fig:letter_PSM_results}(c). The profile of $T_{\text{eff}}$ is similar for the classical and quantum cases and resembles the behavior observed in $e_\infty(e_0)$ for the $\phi^4$ model.
Notably, we find that $r^2\simeq 1$ for both classical and quantum models, indicating that the classical FDT effectively characterizes the relation between $\overline C$ and $\overline R$, even for $\Gamma=1$, {and suggesting that the long-time dynamics of the driven PSM is not significantly affected by quantum fluctuations. This feature, which was already observed in the undriven version of the model~\cite{cugliandolo1999real}, is likely enhanced by energy absorption from the drive.}

{
\emph{Discussion and conclusion} --- In conclusion, we studied the dynamics of two ergodic systems with unbounded phase space, the classical lattice $\phi^4$ model and the classical or quantum $p$-spin spherical model (PSM), under smooth periodic driving. Both models exhibit clear signatures of prethermalization at finite driving frequencies, unlike previously studied cases. We interpreted this prethermalization as resulting from an interplay between parametric resonance and non-linearity, supported by a perturbative analytical framework—the multiple-scale analysis (MSA)—to estimate the prethermal Hamiltonian ${\mathcal H}_F$. 
For the $\phi^4$ model, we showed that ${\mathcal H}_F$ accurately reproduces the dynamics in the limit of small non-linearity and that it does not asymptotically match the time-averaged Hamiltonian $\overline{H(t)}$. Instead, ${\mathcal H}_F$ emerges as a sum of terms linked to resonance and non-linearity
For each model, we showed that finite-frequency prethermalization in many-body systems is a generalization of  the effective conservation law observed in their single-body counterparts, arising from the same interplay of resonance and non-linearity. Given its model-independent origins, this phenomenon likely extends to a broader class of systems with unbounded energy spectra.
Finally, since the MSA can be applied to various models with unbounded phase space, aexploring the relationship between ${\mathcal H}_F$ and $\overline{H(t)}$ is a promising future direction. Such insights could guide the experimental realization of long-lived many-body Hamiltonians, even at moderate driving frequencies.
}

\emph{Acknowledgements ---}  
We thank D. Abanin, A. Delmonte, A. Polkovnikov and A. Russomanno  for discussions. L.C. acknowledges hospitality from the Institute of Physics of Collège de France. L.F.C. acknowledges support from the grant 
ANR-19-CE30-0014. A.S. acknowledges funding from the grants 
PNRR MUR project PE0000023-NQSTI and Quantera project SuperLink. 
M.S. acknowledges funding from the European Research Council (ERC) under the European Union's Horizon 2020 research and innovation programme (Grant agreement No. 101002955 -- CONQUER).

\bibliography{bibliography.bib}

\newpage

\begin{widetext}

\appendix

\section{\Large Supplemental Material: Finite-frequency prethermalization \\
in periodically driven ergodic systems} 

In Section~\ref{sec:SM_Floquet_therm}, we show that the driven dynamics of the classical one-dimensional lattice $\phi^4$ model deviates from its prethermal plateau when integrated at longer times and discuss the dependence of prethermalization on the degree of non-linearity in the system.
In Section~\ref{sec:SM_MSA}, we provide a comprehensive exposition of the multiple-scale analysis, applied to the lattice $\phi^4$ model. In Section~\ref{sec:SM_mode_coupling}, we write down and derive explicitly the mode coupling equations, which govern the dynamics of the correlation and response functions, in the thermodynamic limit of the $p$-spin spherical model.
In Section~\ref{sec:SM_ergodicity}, we discuss the ergodic properties of both the two models, in absence of a time-dependent drive. 
Finally, in Section~\ref{sec:SM_different_drivings}, we study the Floquet dynamics of both models, driven by smooth periodic forces different from the ones considered in the Letter.

\section{Deviations from the prethermal plateaus}\label{sec:SM_Floquet_therm}

{In this section,  we extend the integration of the dynamics of both the lattice $\phi^4$ model and of the $p$-spin spherical model (PSM) over a longer time scale. This allows us to provide evidence of thermalization in their classical dynamics beyond the prethermal time scale investigated in the main text.}

\subsection{Classical lattice \texorpdfstring{$\phi^4$}{} model}

In this subsection, we integrate Eqs.~\eqref{eq:SM_generic_Hamilton_eq} over a longer time scale  to investigate the behavior of the classical lattice $\phi^4$ model. We compute the resulting average energy density $e(t)=\braket{H_{\phi^4}(\textbf{x}(t),\dot{\textbf{x}}(t))}_0/N$, where $H_{\phi^4}(\textbf{x}(t),\textbf{p}(t))$ is defined in Eq.~(1) of the Letter. As in the Letter, the initial condition is chosen as ensemble of microcanonical configurations, sampled from the surface $H_{\phi^4}\big(\mathbf{x}(0),\dot{\mathbf{x}}(0)\big) = N e_0$ for a fixed initial energy density $e_0$.
To integrate up to a larger maximum time, we fix $\mathcal{N}$ to a smaller value then in the Letter. We integrate the dynamics for different values of $\lambda$, the parameter controlling the non-linearity of the system. From Fig.~\ref{fig:phi4_SM_deviations}(a), it is already evident that the energy density $e(t)$ displays deviations from its finite plateau at $e_\infty$ at sufficiently long-time, confirming the prethermal nature of the plateau {and suggesting an eventual thermalization to infinite temperature}. We also observe that the height of the plateau increases by decreasing $\lambda$.\\

To qualitatively assess the influence of the non-linearity on the prethermal plateau, we introduce the normalized quantity
\begin{equation}\label{eq:phi4_SM_resc_energy}
Q(t) = \frac{e(t)-e_0}{e_\infty-e_0} \ ,
\end{equation}
which saturates to $1$ when $e(t)$ saturates to $e_\infty$ and such that $Q(0)=0$.
Here, we estimate $e_\infty$ as $e(t=t_0)$ for a fixed intermediate time $t_0$. As depicted in Fig.~\ref{fig:phi4_SM_deviations}(b), deviations of $Q(t)$ from its plateau appear at earlier times as $\lambda$ increases, suggesting that a stronger non-linearity suppresses prethermalization.

\begin{figure}[h]
    \centering
    \includegraphics[width=\textwidth]{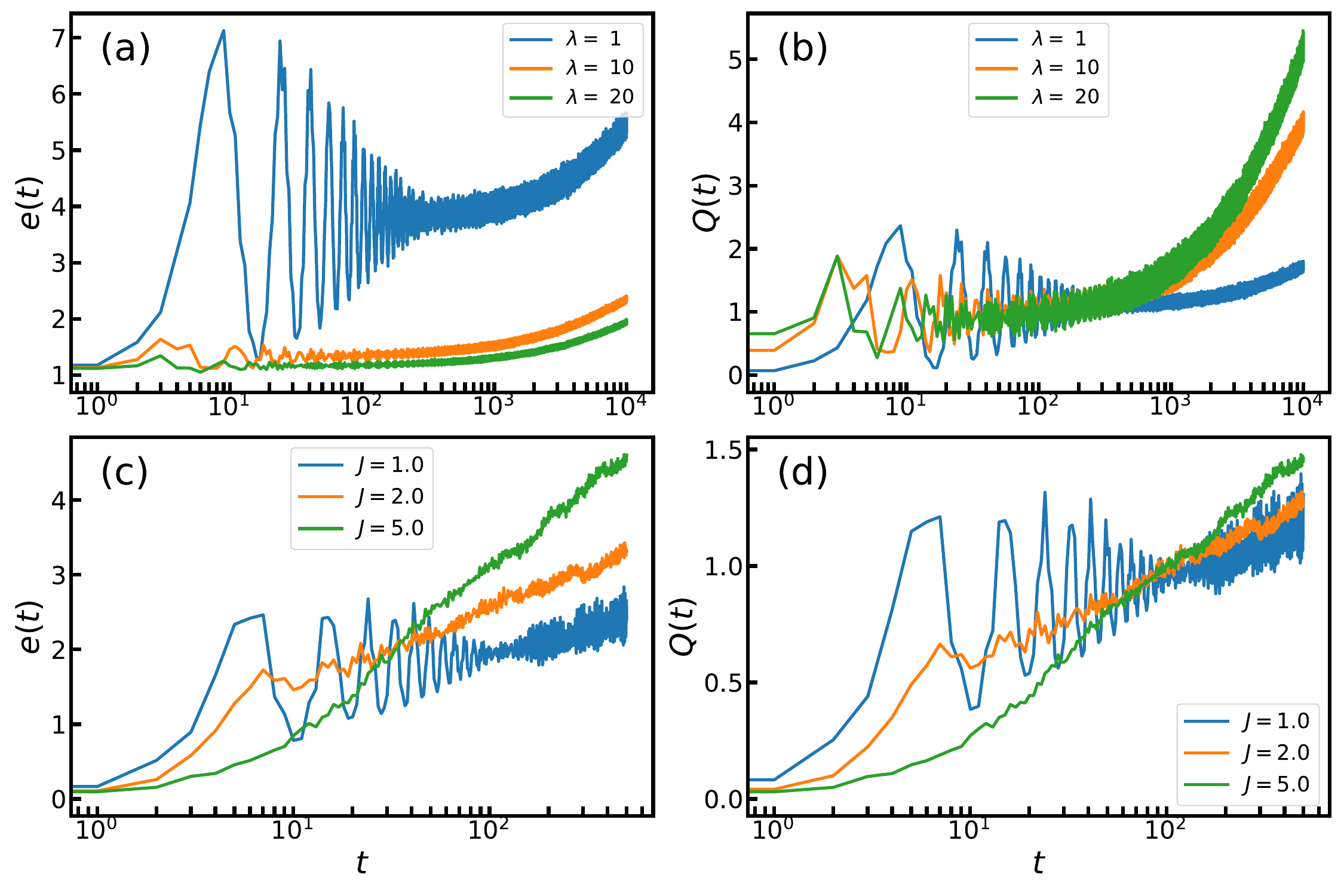}
    \caption{Time evolution of average energy densities, for a driving frequency $\Omega=1.5$. \textbf{(a-b)} One dimensional lattice $\phi^4$ model, from Eqs.~\eqref{eq:SM_generic_Hamilton_eq_gen}, with $N=100$. Data are averaged over $\mathcal{N}=50$ configurations. We set $\omega_0^2=B_0=g=1$ and $e_0=1$. (a)   $e(t)=\braket{H_{\phi^4}(\textbf{x}(t),\textbf{p}(t))}_{0}/N$. (b) $Q(t)$ from Eq.~\eqref{eq:phi4_SM_resc_energy}. We estimate $e_\infty$ as $e(t_0=100)$. \textbf{(c-d)} Analogous plots for the  $p$-spin spherical model, evolving from Eqs.~\eqref{eq:SM_EoM_classical_PSM}, with $N=100$. Data are averaged over $\mathcal{N}=25$ configurations. We set $M=J=B_0=1$, $p=3$ and $e_0=0$. (c) $e(t)=\braket{H_{J}(\boldsymbol{\sigma}(t),\dot{\boldsymbol{\sigma}}(t))}_0/N$. (d) Rescaling $Q(t)$ from Eq.~\eqref{eq:phi4_SM_resc_energy}, with $e_\infty=e(t_0=50)$.}
    
    \label{fig:phi4_SM_deviations}
\end{figure}

\subsection{\texorpdfstring{$p$}{}-spin spherical model}

{In this subsection, we integrate the classical dynamics of the PSM beyond the prethermalization time-scale and for a system with a finite size $N$. The classical equations of motion are given by:
\begin{equation}\label{eq:SM_EoM_classical_PSM}
    M\ddot{\sigma}_i = -z(t)\sigma_i -\sum_{i_2 < ... < i_p} J_{i_1...i_p} \! \sigma_{i_2}...\sigma_{i_p} + B_0\sin(\Omega t) ~ ,
\end{equation}
for $i = 1,\dots,N$. As in the Letter, the  all-to-all couplings $J_{i_1,\dots,i_p}$ are independently sampled from a Gaussian distribution with zero mean and variance $\overline{J^2}=p! J^2/2 N^{p-1}$.
The Lagrange multiplier $z(t)$ implements the spherical constraint, $\sum_{i=1}^N \sigma_i^2 = N$. Given this constraint, it can be easily shown that
\begin{equation}
    z(t)  = \frac{M}{N} \sum_i\dot{\sigma}_i^2-\frac{p}{N}\sum_{i_1 < ... < i_p} J_{i_1...i_p} \! \sigma_{i_1}...\sigma_{i_p} -\frac{B_0\sin(\Omega t)}N \sum_{i=1}^N \sigma_i ~.
\end{equation}
As we did for the lattice $\phi^4$ model, we integrate the dynamics from a set of random initial conditions, sampled uniformly from the constant energy manifold $H_{J}\big(\boldsymbol{\sigma}(0),\dot{\boldsymbol{\sigma}}(0)\big) = N e_0$, where $H_{J}\big(\boldsymbol{\sigma},\dot{\boldsymbol{\sigma}}\big)$ is the classical version of the Hamiltonian from Eq.(13) of the main text. We compute the average classical energy density, $e(t)=\braket{H_{J}(\boldsymbol{\sigma}(t),\dot{\boldsymbol{\sigma}}(t))}_0/N$ and its rescaling from Eq.~\eqref{eq:phi4_SM_resc_energy}. The results, shown in Fig.~\ref{fig:phi4_SM_deviations}(c), indicate that $e(t)$ eventually deviates from the prethermal plateau, potentially thermalizing to infinite temperature, similar to the $\phi^4$ model. Furthermore, the plots of $Q(t)$ in Fig.~\ref{fig:phi4_SM_deviations}(d) show that the length of the plateau increases by decreasing the disorder strength $J$.}

\section{Detailed calculations on multiple-scale analysis}\label{sec:SM_MSA}

In this section, we aim to provide a review on the multiple-scale analysis (MSA) that we employed in the Letter. {To prove its generality, in the first subsection we show its application to an interacting, non-linear set of generic oscillators. Then, in the second subsection, we apply the MSA to the classical and single-body version of the $p$-spin spherical model, which was analyzed in the Letter.}

\subsection{Non-linear oscillators}
{We present its derivation for a generic system of non-linear oscillators, driven by a generic sinusoidal wave. The corresponding equations of motion are
\begin{equation}\label{eq:SM_generic_Hamilton_eq}
    \ddot{x}_i + \omega_0^2 x_i = -\epsilon \frac{\partial V(\mathbf{x})}{\partial x_i} - \epsilon \sin( \Omega t) \frac{\partial U(\mathbf{x})}{\partial x_i}\ ,
\end{equation}
for $i=1\ldots N$. Here, $U(\mathbf{x})$ is driving term of the Hamiltonian, while $V(\mathbf{x})$ contains the interacting and non-linear term of the time-average Hamiltonian. We assume that both $U(\mathbf{x})$ and $V(\mathbf{x})$ are smooth function of $\mathbf{x}$, so that they can be expressed through their Taylor expansion:
\begin{gather}
    U(\mathbf{x}) = \sum_{\mathbf{i}} \Tilde{U}_\mathbf{i}~x_1^{i_1}\cdots x_N^{i_N} \\
    V(\mathbf{x}) = \sum_{\mathbf{i}} \Tilde{V}_\mathbf{i}~x_1^{i_1}\cdots x_N^{i_N}
\end{gather}}
We rescale the harmonic coupling as $\omega_0^2 = \Omega^2 - \epsilon \Delta\omega^2$ and work in the limit of small $\epsilon$, while keeping all other parameters of order one.
As outlined in the Letter, the starting point of the MSA consists in introducing a redundant time variable $\tau = \epsilon t$. $\tau$ defines a long time scale, since it becomes significant only when 
$t$ is of order $\epsilon^{-1}$ or larger. Although we expect that the actual solution $\mathbf{x}(t)$ to Eq.~\eqref{eq:SM_generic_Hamilton_eq} is a function of $t$ alone, MSA seeks solutions which are functions of 
both variables $t$ and $\tau$, treated as independent from each other. Such expression of $\mathbf{x}(t)$ as a function of two variables is an artifice to remove secular effects; the 
actual solution has $t$ and $\tau$ related by  $\tau = \epsilon t$. The formal procedure consists in assuming a perturbative expansion in powers of $\epsilon$:
\begin{equation}
   \mathbf{x}(t) = \mathbf{x}_0(t,\tau) + \epsilon \ \mathbf{x}_1(t,\tau) + \dots \ ,
\end{equation}
for small $\epsilon$.
We use the chain rules $d/dt = \partial_t + \epsilon \partial_\tau$ for partial differentiation to compute the derivatives of the variables $x_i(t)$, obtaining
\begin{equation}\label{eq:SM_first_derivative_multiscale}
    \frac{dx_i}{dt} = \frac{\partial  x_{0,i}}{\partial t} + \epsilon \Big( \frac{\partial x_{1,i}}{\partial t} + \frac{\partial x_{0,i}}{\partial \tau} \Big) + O(\epsilon^2)  
\end{equation}
and
\begin{equation}\label{eq:SM_second_derivative_multiscale}
    \frac{d^2 x_i}{dt^2} = \frac{\partial^2  x_{0,i}}{\partial t^2} + \epsilon \Big( \frac{\partial^2 x_{1,i}}{\partial t^2} + 2 \ \frac{\partial^2 x_{0,i}}{\partial t  \ \partial\tau} \Big) + O(\epsilon^2) \ . 
\end{equation}
By substituting Eq.~\eqref{eq:SM_second_derivative_multiscale} in  Eq.~\eqref{eq:SM_generic_Hamilton_eq} and collecting powers of $\epsilon$, we deduce
\begin{gather} 
    \frac{\partial^2 x_{0,i}}{\partial t^2}  +  \Omega^2 x_{0,i} = 0 \  ,\label{eq:SM_order_0_phi4}\\
    \frac{\partial^2 x_{1,i}}{\partial t^2} + \Omega^2 x_{1,i} = -2 \frac{\partial^2 x_{0,i}}{\partial t \ \partial\tau} + \Delta\omega^2 x_{0,i} -\frac{\partial V(\mathbf{x}_0)}{\partial x_i} -\sin(\Omega t) \frac{\partial U(\mathbf{x}_0)}{\partial x_i} \ . \label{eq:SM_order_1_phi4}
\end{gather}
The general solution to Eq.~\eqref{eq:SM_order_0_phi4} is $ x_{0,i}(t,\tau) = A_i(\tau)e^{i \Omega t} + A_i^*(\tau)e^{-i \Omega t}$. {We aim rewrite the right-hand side of Eq.~\eqref{eq:SM_order_1_phi4} in terms of $A_i$ and $A_i^*$. To do so, we first evaluate the terms $x_j^{i_j}$ and $i_j x_j^{i_j-1}$, which read as
\begin{gather}
   x_j^{i_j} =  \sum_{k=0}^{i_j} \binom{i_j}{k} A_j^k (A^*_j)^{i_j-k} e^{i(2k-i_j)\Omega t}\\
   i_j x_j^{i_j-1} = \sum_{k=0}^{i_j-1} i_j\binom{i_j-1}{k} A_j^k (A^*_j)^{i_j-1-k} e^{i(2k-i_j-1)\Omega t}   = \sum_{k=0}^{i_j} \binom{i_j}{k} A_j^k \frac{\partial }{\partial A_j^*}(A^*_j)^{i_j-k} \ e^{i(2k-i_j-1)\Omega t}  \ .
\end{gather}
Here and in the following, we omit the dependence on $\tau$ to keep the notation compact.
Plugging these expressions into the derivative $\partial V(\mathbf{x}_0)/\partial x_i$ and posing $i=1$ for simplicity, we obtain:
\begin{equation}
    \frac{\partial V(\mathbf{x}_0)}{\partial x_1} = 2 \sum_\mathbf{i}  \Tilde{V}_\mathbf{i}~ {\sum_\mathbf{k}} \binom{i_1}{k_1} A_1^{k_1} \frac{\partial }{\partial A_1^*}(A^*_1)^{i_1-k} \Big[\prod_{j=2}^N \binom{i_j}{k_j} A_j^{k_j}(A_j^*)^{i_j-k_j} \Big] e^{i \sum_{j=1}^N (2k_j-i_j-1) \Omega t} 
\end{equation}}
{A similar result is retrieved for the driving term, proportional to  $\partial U(\mathbf{x}_0)/\partial x_i$. As discussed in the letter, the dynamics of the  slowly varying amplitude $A_i(\tau)$ and its complex conjugate $A_i^*(\tau)$ are determined by imposing that secular terms, defined as solution of the associated homogeneous differential equation, do not appear on the right-hand side of  Eq.~\eqref{eq:SM_order_1_phi4}. In the case under study, these terms are the ones proportional either to $e^{i \Omega t}$ or $e^{-i\Omega t}$.
After some algebra, it can be shown that this requirement is equivalent to imposing the following equation of motion
\begin{equation}
    \frac{dA_i}{d\tau} =  \frac{\partial \mathcal{H}_F}{\partial A_i^*}(\mathbf{A},\mathbf{A}^*) 
\end{equation}
and its complex conjugate. Thus, the dynamics of $A_i$ and $A_i^*$ is Hamiltonian, with an effective Hamiltonian 
\begin{equation} \label{eq:SM_general_eff_ham}
    \mathcal{H}_F(\mathbf{A},\mathbf{A}^*) = -\frac{\Delta\omega^2}{2}\sum_i |A_i|^2 + \sum_{\mathbf{i}} \Tilde{V}_\mathbf{i} ~ {\sum_\mathbf{k}}^\prime \prod_{j=1}^N \binom{i_j}{k_j} A_j^{k_j}(A_j^*)^{i_j-k_j} 
    + \sum_{\mathbf{i}} \Tilde{U}_\mathbf{i} ~ {\sum_\mathbf{k}}^{\prime\prime} \prod_{j=1}^N \binom{i_j}{k_j} A_j^{k_j}(A_j^*)^{i_j-k_j} \ .
\end{equation} 
In Eq.~\eqref{eq:SM_general_eff_ham}, the sum ${\sum_\mathbf{k}}^\prime$ is restricted over the $N$-dimensional indices $\mathbf{k}$ such that $\sum_{j=1}^N 2k_j=i_j+2$, while the sum ${\sum_\mathbf{k}}^{\prime\prime}$ includes the indices either obeying the constraints $\sum_{j=1}^N 2k_j=(i_j+1)$ and $\sum_{j=1}^N 2k_j=(i_j+3)$. Eq.~\eqref{eq:SM_general_eff_ham} can be rewritten in terms of the coordinates $u_i(\tau) = A_i(\tau) + A_i^*(\tau)$ and $v_i(\tau) = [ A_i(\tau) - A_i^*(\tau) ]/i$. The result is a generalization of the effective Hamiltonian presented in the Letter, valid for any many-body Floquet system with unbounded energy spectrum, in the limit of small non-linearity and driving amplitude. }
\subsection{Single-body version of the \texorpdfstring{$p$}{}-spin spherical model}

\begin{figure}[ht]
    \centering
    \includegraphics[width=0.75\textwidth]{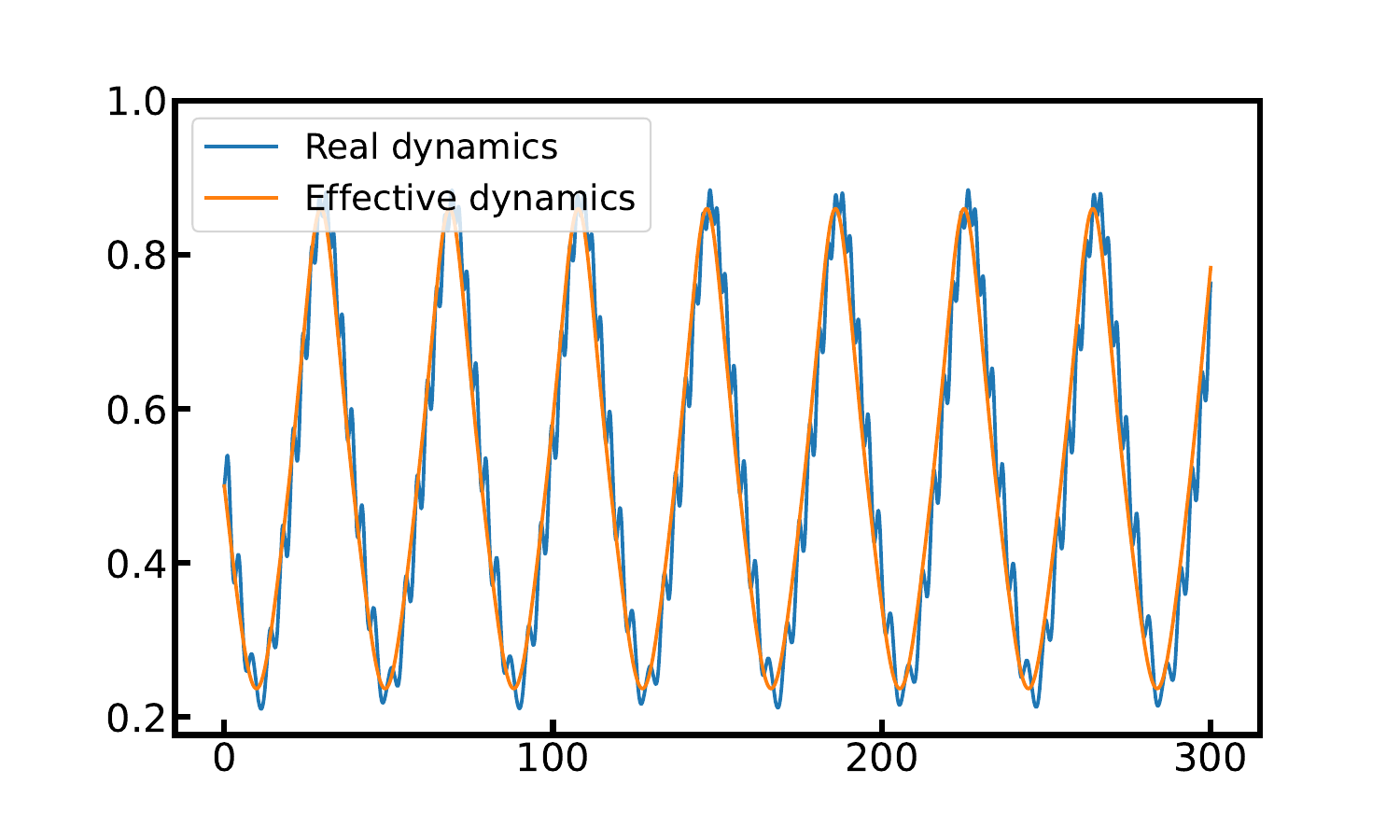}
     \caption{(Color online) Kinetic energy $\dot{\theta}(t)^2/2$ of the $p$-spin spherical model, for $N=p=2$. The blue plot for $\theta(t)$ is obtained by integrating Eq.\,\eqref{eq:SM_pendulum_eq}. The orange plot is obtained by setting $\theta(t)=\alpha_0(\epsilon t) + \Omega t$, where $\alpha_0(\tau)$ is determined by Eq.\,\eqref{eq:SM_alpha_MSA}, with $d\beta_1/d\tau=0$, as discussed in the text. We set an initial condition $\theta(0)=\alpha_0(0)=1$ and $\dot{\theta}(0)=\Omega=1$. Other parameters are $\Tilde{B}_0=\Tilde{J}=1$.}
    \label{fig:SM_pendulum}
\end{figure}

{In this subsection, we apply the MSA to the classical and single-body version of the classical $p$-spin spherical model (PSM). To do so, we study Eqs.~\eqref{eq:SM_EoM_classical_PSM} for $N=2$. We also fix $p=2$, as $p\leq N$ by definition. Taking into account the spherical constraint and introducing the polar angle $\theta$, we write $\sigma_1=\sqrt{2}\cos{\theta}$ and $\sigma_2=\sqrt{2}\sin{\theta}$.
In polar coordinates, it is straightforward to prove that the classical dynamics is determined the following equation of motion:
\begin{equation}\label{eq:SM_pendulum_eq}
    \ddot{\theta} = -2B_0\sin(\Omega t)\sin\left(\theta-\frac{\pi}{4}\right) -2J\cos(2\theta) ~ ,
\end{equation}
To apply a MSA to Eq.~\eqref{eq:SM_pendulum_eq}, we rescale the couplings as $B_0=\epsilon^2\Tilde{B}_0$ and $J=\epsilon^2\Tilde{J}$ and introduce a redundant slow-time variable $\tau=\epsilon t$. The scaling of the couplings with $\epsilon^2$ is chosen empirically. We also set $\dot{\theta}(0)=\Omega$. In the limit of small $\epsilon$, this last choice leads a resonance between the driving frequency $\Omega$ and the frequency of rotations of the system around the circle, which is given by $\dot{\theta}$ and is nearly constant over the 'fast' time scale $t$.
We analyze this system using the MSA. As in the previous subsection, we propose expansion $\theta(t)=\theta_0(t,\tau)+\epsilon \theta_1(t,\tau)+\ldots$. The equation of motion at leading order becomes $\partial_t \theta_0 = 0$ and its solution is $\theta_0(t,\tau) = \alpha_0(\tau) + \beta_0(\tau)t$. The equation at order $\epsilon$ is given by
\begin{equation}\label{eq:SM_pendulum_eps_eq}
    \frac{\partial^2 \theta_1}{\partial t^2} = -2\frac{d \beta_0}{d\tau}
\end{equation}
To remove secularities, we must ensure that all the terms on the right-hand side which depend linearly on the variable $t$ vanish. Thus, we impose $d\beta_0/d\tau=0$. This result, together with the resonant initial condition $\dot{\theta}(0)=\Omega$, implies that $\beta_0(\tau)=\beta_0(0)=\Omega$. At the same time, we obtain that $\theta_1(t,\tau) = \alpha_1(\tau) + \beta_1(\tau)t$ is the solution of Eq.~\eqref{eq:SM_pendulum_eps_eq}.\\
At this point, we have fully determined $\beta_0(\tau)$, while $\alpha_0(\tau)$ is still unknown. To compute $\alpha_0$, we investigate the equation at order $\epsilon^2$, given by
\begin{equation}
    \frac{\partial^2 \theta_2}{\partial t^2} = -2\frac{d\beta_1}{d\tau} -\frac{d^2\alpha_0}{d\tau^2} -2\Tilde{J} \cos(2\alpha_0+2\Omega t)+ \Tilde{B}_0 \cos \left(2\Omega t + \alpha_0-\frac{\pi}{4}\right) - \Tilde{B}_0\cos\left(\alpha_0-\frac{\pi}{4}\right)
\end{equation}
By removing secularities, we are lead to the equation 
\begin{equation}\label{eq:SM_alpha_MSA}
    \frac{d^2\alpha_0}{d\tau^2} = -2\frac{d\beta_1}{d\tau} - \Tilde{B}_0\cos\left(\alpha_0-\frac{\pi}{4}\right) ~ .
\end{equation}
The function $\beta_1'(\tau) = d\beta_1(\tau)/d\tau$ can be determined by the equation at order $\epsilon^3$. For brevity, we refrain from performing the calculation in detail and leave it to the reader. Instead, we only mention that, by eliminating secular terms at order $\epsilon^3$, it can be shown that $\beta_1^\prime(\tau)=\beta_1^\prime(0)$ is a constant.
However $\beta_1^\prime(0)$ remains undetermined from our calculation, though it is possibly fixed by even higher perturbative orders. Here, we fix $\beta_1^\prime(0)=0$ empirically.
With this choice, the MSA reproduces the original dynamics faithfully for small $\epsilon$, as shown in Fig.~\ref{fig:SM_pendulum}. This result further demonstrates the generality of the MSA, as the calculation reported in this subsection can be straightforwardly generalized to many-body systems with degrees of freedom defined over a compact manifold.
}

\section{Dynamical equations for the correlation and response function in the \texorpdfstring{$p$}{}-spin spherical model}\label{sec:SM_mode_coupling}

The aim of this section is to derive and discuss the closed set of equations of motions governing the dynamics of the correlation and response functions, $C(t,t')$ and $R(t,t')$, of the $p$-spin spherical model (PSM) investigated in the Letter.
We anticipate that, in analogy to the results obtained in Refs.~\cite{Berthier2001,Busiello2004drivenPSM}, which studied the periodically driven dynamics of the PSM coupled to a thermal bath, the equations describing its Floquet dynamics have Schwinwger-Dyson structure and are given by
\begin{align}
[ M \partial_t^2 + z(t)] R(t,t') &= \delta(t-t') + \int_{0}^{t} dt'' \; \Sigma(t,t'') R(t'',t') \label{eq:SM_mc_resp}
\; , 
\nonumber \\
[ M \partial_t^2 + z(t)]  C(t,t') &= \int_{0}^{t} 
dt'' \;  \Sigma(t,t'')C(t'',t') + \int_0^{t'}  dt'' \; D(t,t'') R(t',t'') \ + 
\\
&+ B(t)\int_0^{t'} dt'' \; R(t',t'')B(t'') \; .
\label{eq:SM_mc_corr}
\end{align}
We will define the self-energy $\Sigma(t,t')$ and the vertex $D(t,t')$  later on in this section.
The time-dependent Lagrange multiplier $z(t)$ enforces the spherical constraint and has an expression in terms of $C$ 
and $R$ as well.
In the Letter, we integrate numerically Eqs.~\eqref{eq:SM_mc_resp} and~\eqref{eq:SM_mc_corr} to compute the correlation and response function, evolving from  the equal-time conditions:
\begin{align}
C(t,t) &= 1 \ , \quad \quad \quad \quad \partial_t C(t,t') |_{t' \to t^-}  = \partial_t C(t,t') |_{t' \to t^+}  = 0 \ , \\ \label{eq:equal_time_conditions}
R(t,t) &= 0 \ , \quad \quad \quad \quad \partial_t R(t,t') |_{t' \to t^-} = \frac{1}{M} \ , \quad \quad \quad \quad \partial_t R(t,t') |_{t' \to t^+} = 0 \ ,
\end{align}
resulting from the spherical constraint and the causal structure of the response function.
Then, following Refs.~\cite{thomson2020quantum,Cugliandolo_2017}, we compute the energy density of the system as
\begin{equation}\label{eq:SM_mc_energy}
    e(t) = -\frac M 2 \partial_t^2 C(t,t')\big|_{t'\to t^-} - \frac{1}{p} \int_0^t dt''  \; \left[ \Sigma(t,t'')C(t,t'') + D(t,t'')R(t,t'') \right] \ .
\end{equation}

This section is organized as follows. In Section~\ref{subsec:SM_derivation_mode_coupling}, we derive Eqs.~\eqref{eq:SM_mc_resp} and~\eqref{eq:SM_mc_corr} using the Schwinger-Keldysh formalism in the large $N$ limit. In Section~\ref{subsec:SM_lagrange_mult}, we show how to determine the Lagrange multiplier $z(t)$ explicitly, in terms of the correlation and response functions. Finally, in Section~\ref{subsec:SM_predictor_corrector}, we discuss the predictor-corrector algorithm, that we use to integrate Eqs.~\eqref{eq:SM_mc_resp} and~\eqref{eq:SM_mc_corr} numerically.

\subsection{Derivation of the Schwinger-Dyson equations}
\label{subsec:SM_derivation_mode_coupling}

We begin by reminding that the Hamiltonian of the driven quantum PSM is defined as
\begin{equation}
\label{eq:SM_driven_hamiltonian_SG}
    \hat{H}_J = \frac 1{2M}\sum_{i=1}^N \ \hat{\Pi}_i^2 -\mathcal{J}(t)\sum_{i_1<\dots<i_p}J_{i_1,\dots,i_p} \hat{\sigma}_{i_1}\cdots \hat{\sigma}_{i_p} - B(t)\sum_i \hat{\sigma}_i
\end{equation}
with all-to-all couplings $J_{i_1,\dots,i_p}$, independently sampled from a Gaussian distribution with zero mean and variance $\overline{J^2}=2J^2 p!/N^{p-1}$.  The spins $\hat{\sigma}_i$ obey the usual spherical constraint $\sum_i \braket{\hat{\sigma}_i^2}=N$ (on average). The canonical commutation relations $[\hat{\sigma}_j,\hat{\Pi}_k]=i\hbar\delta_{jk}$ hold. 
We added a time-dependent factor ${\mathcal J}(t)$
to allow for other kinds of drives.

The unitary dynamics generated by the Hamiltonian in Eq.~\eqref{eq:SM_driven_hamiltonian_SG} can be expressed through a path integral on the Schwinger-Keldysh contour~\cite{kamenev2011field}, whose generating functional reads
\begin{align}
\mathcal{Z}[
J] &= \int \mathcal{D}\boldsymbol{\sigma}^+ \mathcal{D} \boldsymbol{\sigma}^- \exp \left[ i \left( S[\boldsymbol{\sigma}^+] - S[\boldsymbol{\sigma}^-]  
\right)/\hbar \right] \braket{\boldsymbol{\sigma}^+| \hat{\rho}(0) |\boldsymbol{\sigma}^-} \ .
\end{align}
Here, $\hat{\rho}(0)$ represents the element of the initial density matrix at $t=0$ and is chosen to be a random infinite-temperature initial state, which is uncorrelated with the disorder. The last factor in the path integral is then just an irrelevant factor. 
The action $S$ can be defined in terms of a quadratic term $S_0$ and a disordered interaction term $V_J(\boldsymbol{\sigma})$ as follows:
\begin{align}
S [\boldsymbol{\sigma}, J] &= S_0[\boldsymbol{\sigma}]  - \int_0^{\infty} dt \; \phantom{.} \mathcal{J}(t) V_J\big(\boldsymbol{\sigma}(t)\big), \\
S_0[\boldsymbol{\sigma}]  &= \int_0^{\infty} dt \; \left[ \frac{M}{2} \dot{\boldsymbol{\sigma}}^{2} - \frac{z(t)}{2} (\boldsymbol{\sigma}^2 - N)\right]- \sum_{i=1}^N \int_0^\infty dt \; B(t)\sigma_i(t), \\
V_J(\boldsymbol{\sigma}) &= -\sum_{i_1 < ... < i_p}^{N} J_{i_1...i_p} \sigma_{i_1}...\sigma_{i_p}.
\end{align}
The quadratic part $S_0$ contains a kinetic term, chosen such that the eventual dynamical equations are written in terms of second derivatives with respect to time. The time-dependent Lagrange multiplier $z(t)$ enforces the constraint $\sum_{i}^{N} \sigma_i^2 = N$. We also included in $S_0$ the coupling to the time-dependent field $B(t)$. 
This action can be split into components $\boldsymbol{\sigma}^{+}$ and $\boldsymbol{\sigma}^{-}$ residing on the forward and backwards Keldysh contours respectively, and 
recast as
\begin{align}
S[\boldsymbol{\sigma}^{+},\boldsymbol{\sigma}^{-}, J]  &= S_0[\boldsymbol{\sigma}^{+}] - S_0[\boldsymbol{\sigma}^{-}] - \int_0^{\infty} dt \;  \mathcal{J}(t) \left[ V_J\big(\boldsymbol{\sigma}^{+}(t)\big) - V_J\big(\boldsymbol{\sigma}^{-}(t)\big) \right]
\end{align}
where the relative minus sign comes from reversing the integration limits on the reverse contour.

\subsubsection{System-Bath Coupling}

The coupling between the system and bath can be treated exactly as in Ref.~\cite{cugliandolo1999real}. 
We couple the quantum $p$-spin Hamiltonian linearly to a bath of harmonic oscilllators, assumed to be in thermal equilibrium, for a time window $[0,t_b]$. This coupling can be described by the functional~\cite{kamenev2011field}:
\begin{align}
S_{bath} = \int_0^{t_b} dt \int_0^{t_b} dt' \; \big\{ -[\boldsymbol{\sigma}^+(t)-\boldsymbol{\sigma}^-(t)] \eta(t-t') [\boldsymbol{\sigma}^+(t') + \boldsymbol{\sigma}^-(t')] \\
+ i ~ [\boldsymbol{\sigma}^+(t) - \boldsymbol{\sigma}^-(t)] \nu(t-t')[\boldsymbol{\sigma}^+(t') - \boldsymbol{\sigma}^-(t')] ~ \big\} \nonumber
\end{align}
where $\eta$ and $\nu$ are the correlation and response functions of the bath, and are time-translation invariant due to the bath being in equilibrium. They are given by
\begin{align}
\eta(t-t') &= -\Theta(t-t') \int_0^{\infty} d \omega I(\omega) \sin [\omega(t-t')]
\; , \\
\nu(t-t') &= \int_0^{\infty} I(\omega) \coth \left( \frac12 \beta \hbar \omega \right) \cos[\omega(t-t')]
\; , 
\end{align}
where $I(\omega)$ is the spectral function of the bath. We choose an Ohmic bath with $I(\omega) = \frac{1}{\pi} \exp \left( -|\omega|/\Lambda \right)$ and set the integration cutoff to be $\Lambda=5$. 

\subsubsection{Disorder Averaging}

For an initial condition $\hat{\rho}(0)$ uncorrelated with the disorder, we can perform the disorder average explicitly. As in the Letter, we assume that the distribution of the disorder variable $J_{i_1...i_p}$ is given by the Gaussian distribution
\begin{align}
P[J] &= \sqrt{\frac{N^{p-1}}{ \pi p! J^2}} \exp \left( - \frac{N^{p-1}}{p! J^2} \sum_{i_1 \neq ... \neq i_p} (J_{i_1...i_p})^{2} \right) \label{Eq.pdist}
\end{align}
with zero average and variance $\overline{(J_{i_1...i_p})^{2} } = (p!J^2)/(2 N^{p-1})$.
The disorder average reduces to just averaging over the terms including the $V_J(\boldsymbol{\sigma})$, which are the only ones depending on the disorder. By an straightforward calculation, we obtain the disorder-averaged generating functional:
\begin{align} 
&\overline{ \mathcal{Z}[
J]} = \int \mathcal{D}{\sigma}^{-} \mathcal{D}\sigma^+ \exp\left[ i \left( S_{eff}[\boldsymbol{\sigma^{+}},\boldsymbol{\sigma^{-}}]
\right)/\hbar \right] \ .
\end{align}
The effective action at the exponent is given by: 
\begin{align}
S_{\rm eff}[\boldsymbol{\sigma^{+}},\boldsymbol{\sigma^{-}}]  & = S_0[\boldsymbol{\sigma}^{+}] - S_0[\boldsymbol{\sigma}^{-}] - V_{D}[\boldsymbol{\sigma}^{+},\boldsymbol{\sigma}^{-}] +S_{bath}[\boldsymbol{\sigma^{+}},\boldsymbol{\sigma^{-}}], \\
V_{D}[\boldsymbol{\sigma}^{+},\boldsymbol{\sigma}^{-}] 
& 
= \frac{i N}{4}  \int dt dt' \; 
{ \mathcal{J}(t) \mathcal{J}(t') J^2} \sum_{\alpha,\beta = \pm}\sum_{i=1}^N \alpha \beta \left( \frac{1}{N}\sigma_i^{\alpha} (t)\sigma_i^{\beta}(t') \right)^{p} 
\end{align}
where $\alpha,\beta=\pm$ are the Schwinger-Keldysh contours.

\subsubsection{Transformed Order Parameters}

The contribution to the action containing at most quadratic terms in the spins $\sigma_i^{\alpha}(t)$ can be written down as:
\begin{align}\label{eq:SM_quadratic_action}
\frac{1}{\hbar} S_{\rm eff}^{(2)}[\boldsymbol{\sigma}^{+},\boldsymbol{\sigma}^{-}] &= -\frac12 \sum_{i,\alpha,\beta} \int dt  dt' \; \sigma_i^{\alpha}(t) O_{p}^{\alpha \beta}(t,t') \sigma_i^{\beta}(t') - \sum_{i,\alpha}\int \frac{dt}{\hbar} \ \alpha B(t) \sigma_i^\alpha(t) \ ,
\end{align}
where again $\alpha,\beta = \pm$.
The matrix elements $O^{\alpha\beta}_p(t,t')$, appearing in Eq.~\eqref{eq:SM_quadratic_action}, are explicitly given by
\begin{align}
O_{p}^{++}(t,t') &= \frac{1}{\hbar}[M \partial_t^{2} + z^{+}(t)] \delta(t-t') - \frac{2}{\hbar} (i\nu(t-t')-\eta(t-t')) \Theta(t-t_b) \Theta(t'-t_b), \\
O_{p}^{+-}(t,t') &= \frac{1}{\hbar} (2 \eta(t-t') + 2 i \nu(t-t'))\Theta(t-t_b) \Theta(t'-t_b), \\
O_{p}^{-+}(t,t') &= \frac{1}{\hbar} (-2 \eta(t-t') + 2 i \nu(t-t'))\Theta(t-t_b) \Theta(t'-t_b), \\
O_{p}^{--}(t,t') &= -\frac{1}{\hbar}[M \partial_t^{2} + z^{-}(t)] \delta(t-t')- \frac{2}{\hbar} (i\nu(t-t')+ \eta(t-t'))\Theta(t-t_b) \Theta(t'-t_b).
\end{align} 

\subsubsection{Saddle-Point Equations}

To decouple the $p$-body interaction term, we introduce a new set of variables $Q^{\alpha \beta}(t,t')$, by using the following trivial representation of the number $1$:
\begin{align}
1 &= \int \prod_{\alpha \beta} \mathcal{D} Q^{\alpha \beta} \delta \left( \frac{1}{N} \boldsymbol{\sigma}^{\alpha}(t) \boldsymbol{\sigma}^{\beta}(t') - Q^{\alpha \beta}(t,t')\right), \\ 
& \propto \int \prod_{\alpha \beta} \mathcal{D} Q^{\alpha \beta} \mathcal{D} \lambda^{\alpha \beta} \exp \left( -\frac{i}{2} \lambda^{\alpha \beta} \left( \boldsymbol{\sigma}^{\alpha}(t) \boldsymbol{\sigma}^{\beta}(t') - N Q^{\alpha \beta}(t,t')\right) \right).
\end{align}
Then, using a compact notation, the final form of the generating functional is
\begin{equation}
    \overline{ \mathcal{Z}[
    J]} = \int \mathcal{D}Q \mathcal{D}\lambda \; \exp\{ N S[Q,\lambda] \}
\end{equation}
where
\begin{equation}
S[Q,\lambda] =  \sum_{\alpha\beta}\int dtdt' \Big\{\frac i 2 \lambda^{\alpha\beta}(t,t')Q^{\alpha\beta}(t,t')- \frac 14 Q_{\alpha\beta}(t,t') \Big\} + \log Z[Q,\lambda]
\end{equation}
and
\begin{equation}
\begin{split}
    Z[Q,\lambda] = \int \mathcal{D}{\sigma}^{-} \mathcal{D}\sigma^+ \exp \Big\{ &-\frac12 \sum_{\alpha\beta}\int dt  dt'\sigma^{\alpha}(t)[iO_{p}^{\alpha \beta}(t,t')+i\lambda^{\alpha \beta}(t,t')]\sigma^{\beta}(t') + \\
    &- i\sum_{\alpha}\int \frac{dt}{\hbar} \alpha B(t) \sigma_i^\alpha(t) \Big\}
\end{split}
\end{equation}
is the action of a single effective spin.
Then, defining the matrix $M^{\alpha \beta}(t,t') = iO_{p}^{\alpha \beta}(t,t')+i\lambda^{\alpha \beta}(t,t')$, the saddle point equations for the $N\to\infty$ limit read
\begin{align}
&Q^{\alpha \beta}(t,t') = \braket{\sigma^{\alpha} (t)\sigma^{\beta}(t')} = (M^{-1})^{\alpha \beta}(t,t') + m(t)m(t')\label{eq:saddle_Q}\\
&i\lambda^{\alpha \beta}(t,t') = \frac{p}{2}F[Q]^{\alpha \beta}(t,t') \label{eq:saddle_lambda}
\end{align}
The average $\braket{\dots}$ in Eq.~\eqref{eq:saddle_Q} is performed over the partition function $Z[Q,\lambda]$ and the magnetization,
\begin{equation}\label{eq:magnetization}
    m(t) = \braket{\sigma(t)} = \sum_{\beta} \int dt' \; \frac{i}{\hbar}  (M^{-1})^{\alpha \beta}(t,t') \beta B(t')\ ,
\end{equation}
does not  actually depend on the Keldysh index $\alpha$, like every other one-time quantity. The matrices used in Eq.~\eqref{eq:saddle_Q} are defined as
\begin{align}
\mathcal{Q}(t,t') &= \begin{bmatrix} Q^{++}(t,t') & Q^{+-} (t,t')\\ Q^{-+}(t,t') & Q^{--}(t,t') \end{bmatrix}, \quad \mathcal{M}(t,t') = \begin{bmatrix} M^{++}(t,t') & M^{+-} (t,t')\\ M^{-+}(t,t') & M^{--}(t,t') \end{bmatrix} \\
F[Q](t,t') &= \begin{bmatrix} [Q^{++} (t,t')]^{p-1} & -[Q^{+-} (t,t')]^{p-1} \\ -[Q^{-+} (t,t')]^{p-1} & [Q^{--} (t,t')]^{p-1} \end{bmatrix}  \frac{\mathcal{J}(t) \mathcal{J}(t') J^2}{\hbar^2}
\end{align}
We used the same notation of Ref.~\cite{cugliandolo1999real}. Manipulating Eq.~\eqref{eq:saddle_Q}, we can rewrite the response function as
\begin{equation}
    R(t,t') =  \frac{i}{\hbar}[Q^{++}(t,t')-Q^{+-}(t,t')] = \frac{i}{\hbar}[(M^{-1})^{++}(t,t')-(M^{-1})^{+-}(t,t')] \ .
\end{equation}
Then, by fixing $\alpha=+1$ in Eq.~\eqref{eq:magnetization}, the magnetization becomes
\begin{equation}
    m(t) = \int dt' \; R(t,t')B(t') \ .
\end{equation}
Using Eq.~\eqref{eq:saddle_lambda} and applying the matrix $\mathcal{M}$ on both sides of Eq. Eq.~\eqref{eq:saddle_Q}, we obtain the dynamical equations for all the two-point correlators on the Keldysh contour, represented in the following compact form:
\begin{align}
i O_p \otimes \mathcal{Q} (t,t')=  \mathcal{I} - \frac{p}{2} F[Q] \otimes \mathcal{Q}(t,t') + \mathcal{S} \ B(t) \braket{\sigma(t')} \label{eq:sp}
\end{align}
where the matrix elements $\mathcal{S}^{\alpha\beta}= \alpha$ do not depend on the time indices. 

\subsubsection{Dynamical Equations}

Following the same prescription described in Ref.~\cite{cugliandolo1999real}, it is straightforward to obtain a closed set of dynamical equations for the response function $R(t,t')$ and the (symmetric) correlation function
\begin{align}
C(t,t') = \frac12 [Q^{+-}(t,t') + Q^{-+}(t,t')] \ .
\end{align}
In particular, the equations of motion for the response function $R(t,t')$ are obtained by taking the difference of the $++$ and $+-$ components of Eq.~\eqref{eq:sp}, while the ones for $C(t,t')$ are obtained by taking the addition of the $+-$ and $-+$ components. After some algebra, the result is given by:
\begin{align}
[ M \partial_t^2 + z(t)] R(t,t') &= \delta(t-t') + \int_{0}^{t} dt'' \; \Sigma(t,t'') R(t'',t') \ , \\
[ M \partial_t^2 + z(t)]  C(t,t') &= \int_{0}^{t} dt'' \; \Sigma(t,t'')C(t'',t') + \int_0^{t'}  dt'' \; D(t,t'') R(t',t'') + \ ,\\
&+ B(t)\int_0^{t'} dt'' \; R(t',t'')B(t'') \ .\nonumber
\end{align}
Here, we have defined the self-energy $\Sigma(t,t')$ and the vertex $D(t,t')$ as
\begin{align}
\Sigma(t,t') = &  -4 \eta(t-t') \Theta(t_b-t)\Theta(t_b-t') \ + \label{eq:SM_PSM_Sigma} \\
& -\frac{p \mathcal{J}(t) \mathcal{J}(t') J^2}{\hbar} \textrm{Im} \left[ C(t,t') - \frac{i \hbar}{2} R(t,t') \right]^{p-1}, \nonumber \\
D(t,t') = & \ 2 \hbar \nu(t-t') \Theta(t_b-t)\Theta(t_b-t') \ + \label{eq:SM_PSM_D} \\
&+ \frac{p \mathcal{J}(t) \mathcal{J}(t') J^2}{2} \textrm{Re} \left[ C(t,t')-\frac{i}{2}(\hbar R(t,t') + \hbar R(t',t)) \right]^{p-1}. \nonumber
\end{align}
This result is exactly the one that we anticipated at the beginning of this section, now with explicit functional forms for the two kernels.

The equations for the classical PSM can be readily obtained from the ones above by simply taking the $\hbar\to 0$ limit. The global structure of the equations is the same, only the $\Sigma$ and $D$ kernels are affected and considerably simplified.

\subsection{The evolution equation for the Lagrange multiplier}\label{subsec:SM_lagrange_mult}

In this section we discuss how to self-consistently determine the Lagrange multiplier $z(t)$.
The approach typically used in the literature~\cite{cugliandolo1999real} consists in evaluating Eqs.~\eqref{eq:SM_mc_resp} and~\eqref{eq:SM_mc_corr}  at equal times $t'=t$:
\begin{equation}\label{eq:constraint_literature}
    z(t) = \int_0^{t} dt'' 
\; 
\Big[ \Sigma(t,t'')C(t'',t)+D(t,t'')R(t,t'')\Big] - M\partial_t^2 C(t,t')\big|_{t'\to t} \ .
\end{equation}
We observe that Eq.~\eqref{eq:constraint_literature} leads to ambiguities, as it tautologically depends on the second derivative of $C(t,t')$ at equal times. 
This issue is irrelevant while the system is coupled to a thermal bath~\cite{cugliandolo1999real}, which naturally leads $z(t)$ to thermalize, or when the total energy is conserved, as the second derivative can be replaced by a causal expression containing the conserved energy density~\cite{thomson2020quantum,Cugliandolo_2017}. However, for $t>t_b$ none of these two condition is met in our protocol, as the system evolves under a unitary dynamics generated by a time-dependent Hamiltonian. In this case, Eq.~\eqref{eq:constraint_literature} does not determine $z(t)$.
Here we solve this issue by determining $z(t)$ from the spherical constraint, according to the following procedure.
We take the total derivative of the constraint equation $C(t,t)=1$ multiple times and obtain:
\begin{equation}\label{eq:constraint_derivatives}
\begin{split}
    0 & = \frac{d}{dt}  C(t,t) \ = \lim_{t'\to t} \partial_tC(t,t') 
    \; , 
    \\
    0 & = \frac{d^2}{dt^2}  C(t,t) = \lim_{t'\to t} \ [\partial^2_t C(t,t') + \partial_t \partial_{t'} C(t,t')] \; , 
    \\
    0 & = \frac{d^3}{dt^3}  C(t,t)  = \lim_{t'\to t} \ [\partial^3_t C(t,t') + 3\partial^2_t \partial_{t'} C(t,t')]
    \; . \\
\end{split}
\end{equation}
We used a compact notation
$\partial_t  = d/dt$ and 
we also used the symmetry relation $C(t,t')=C(t',t)$ in the last equality of each line of Eqs.~\eqref{eq:constraint_derivatives}. It is easy to realize that the first line of Eqs.~\eqref{eq:constraint_derivatives} corresponds to some of the equal-time conditions described in Eqs.~\eqref{eq:equal_time_conditions} and that the second line is useless, as $\partial_t \partial_{t'} C(t,t')$ cannot be computed using the dynamical equations. 
The third line is instead the one that we use to determine $z(t)$. Specifically, we observe that the right-hand-side of its second equality can be rewritten in terms of partial derivatives, performed with respect to $t$ and $t'$ respectively, of both sides of Eq.~\eqref{eq:SM_mc_corr}. After some algebra, we obtain
\begin{align}\label{eq:multiplier_eq}
    \frac{dz}{dt} &= \int_0^{t} dt'' 
    \; \Big[
    \partial_t \Sigma(t,t'')C(t,t'')+ \partial_t D(t,t'')R(t,t'') + 3 \Sigma(t,t'') \partial_t C(t,t'')+ 3 D(t,t'') \partial_t R(t,t'') \Big] \ + \nonumber \\ 
    &+ \partial_t B(t) \int_0^t dt'' \; R(t,t'') B(t'') + 3 B(t) \int_0^t dt'' \; \partial_t R(t,t'') B(t'') 
\end{align}
From Eq.~\eqref{eq:multiplier_eq} it is easy to determine the Lagrange multiplier in a causal form, as
\begin{align}\label{eq:multiplier_integral}
    z(t) &= \int_{t_0}^t dt' \int_0^{t'} dt'' \; \Big[ \partial_{t'} \Sigma(t',t'')C(t,t'')+ \partial_{t'} D(t',t'')R(t',t'')  + 3 \Sigma(t',t'') \partial_{t'} C(t',t'')+ 3 D(t',t'') \partial_{t'} R(t',t'') \Big] + \nonumber \\
    & \ + \int_{t_0}^t dt' \int_0^{t'} dt'' \; \Big[ \partial_{t'} B(t') R(t,t'') B(t'') + 3 B(s) \partial_{t'} R(t',t'') B(t'') \Big] + z(t_0)
\end{align}
provided that we can access the value $z(t_0)$, for some $t_0$. Thus we first solve numerically the dissipative dynamics generated from Eqs.~\eqref{eq:SM_mc_resp} and \eqref{eq:SM_mc_corr}, for $0<t<t_b$, using the standard expression from Eq.~\eqref{eq:constraint_literature}. Subsequently, we solve the ``closed dynamics'' for $t>t_b$ using the expression from Eq.~\eqref{eq:multiplier_integral} for $t_0=t_b$ and by using the knowledge of $z(t_b)$ that we obtained by integrating the dissipative dynamics.\\

\subsection{Predictor-corrector scheme for the Mode-Coupling equations}\label{subsec:SM_predictor_corrector}

To solve Eqs.~\eqref{eq:SM_mc_resp} and~\eqref{eq:SM_mc_corr} in the Letter, we first introduce a discrete time step $\Delta t$ and we discretize the times  as $t=n\Delta t$, $t'=m\Delta t$, etc. for $n$ and $m$ non-negative integers. We use a ``forward'' discretization scheme for the time derivatives, that is
\begin{equation}
    \partial_t f(t,t') \approx (f_{n+1,m} - f_{n,m})/\Delta t \ ,
\end{equation}
for every two-point function $f(t,t')$.
With this notation, the discretized version of  Eqs.~\eqref{eq:SM_mc_resp} and~\eqref{eq:SM_mc_corr} equations is
\begin{equation} \label{eq:quantum_MC_discrete}
\left\{
    \begin{split}
    R_{n+1,m} &= R_{n,m} + \Pi^{(R)}_{n,m}\Delta t/M 
    \; , 
    \\ 
    C_{n+1,m} &= C_{n,m} + \Pi^{(C)}_{n,m} \Delta t/M
    \; , 
    \\ 
    \Pi^{(R)}_{n+1,m} &= \Pi^{(R)}_{n,m} -z_n \Delta t R_{n,m} + \delta_{n,m} + \Delta t F^{(1)}_{n,m} 
    \; , 
    \\
    \Pi^{(C)}_{n+1,m} &= \Pi^{(C)}_{n,m}-z_n\Delta t C_{n,m} + \Delta t F^{(2)}_{n,m} 
    \; , 
    \nonumber
    \end{split}
\right.
\end{equation}
where
{
\begin{equation}
\left\{
    \begin{split}
    F^{(1)}_{n,m} & \ = \ \Delta t \sum_{j=m}^n \Sigma_{nj}R_{jm} 
    \; , 
    \\
    F^{(2)}_{n,m}  & \ = \  \Delta t \ \Big[ \sum_{j=0}^m D_{nj}R_{jm} + \sum_{j=0}^n \Sigma_{nj}C_{jm} \Big] 
    \; , 
    \\
    \Sigma_{n,m} & \ = \   -\frac {p \mathcal{J}(n\Delta t) \mathcal{J}(m\Delta t) J^2} \hbar \ \text{Im}\Big[C_{n,m}- \frac{i\hbar}2 R_{n,m} \Big]^{p-1}~ - \ 4\eta_{n,m}\Theta(t_b-n\Delta t)\Theta(t_b-m\Delta t)
    \; , 
    \\
    D_{n,m}  & \ = \  \frac {p \mathcal{J}(n\Delta t) \mathcal{J}(m\Delta t) J^2} \hbar \ \text{Re}\Big[C_{n,m}- \frac{i\hbar}2 \big(R_{n,m}+R_{m,n}) \Big]^{p-1} ~ + \ 2\hbar\nu_{n,m}\Theta(t_b-n\Delta t)\Theta(t_b-m\Delta t)
    \; .
    \end{split}
\right.
\end{equation}}
The equal time conditions, from Eqs.~\eqref{eq:equal_time_conditions}, here assume the simple form
\begin{equation}
    C_{n,n}=0 \ , \qquad\qquad\quad R_{n,n} = \Pi^{(R)}_{n,n} = \Pi^{(C)}_{n,n} = 0 \ .
\end{equation}
It is easy to note that Eqs.~\eqref{eq:quantum_MC_discrete} are causal, implying that we can iteratively compute each minor matrix $C_{0\leq i\leq n,0\leq j\leq n}$ from the knowledge of $C_{0\leq i\leq n-1,0\leq j\leq n-1}$. To understand in detail why, we notice that the row $C_{n,0\leq j \leq n-1}$ can be straightforwardly computed from the minor $C_{0\leq i\leq n-1,0\leq j\leq n-1}$ using Eqs.~\eqref{eq:quantum_MC_discrete}, then the column $C_{0\leq i\leq n-1,n}$ is immediately obtained from the symmetry relation $C_{n,m}=C_{m,n}$ and the diagonal element $C_{n,n}=1$ is given by the spherical constraint. The same reasoning holds for the response function, with the only difference that $R_{n,m}$ is not symmetric and  $R_{0\leq i\leq n-1,n}=0$ due to causality.\\

We aim to integrate Eqs.~\eqref{eq:SM_mc_resp} and~\eqref{eq:SM_mc_corr} up to some finite time $t_{max}=n_{steps}\Delta t$. With the discretization we chose, the error we make is of order $n_{steps}\Delta t^2$. In order to increase the precision of our results, we improve our method by employing a predictor-corrector algorithm, already used in a similar context in Ref.~\cite{Haldar2020quench}. In a nutshell, we first predict the $(n+1)$-th row of the correlation and response function using Eq.~\eqref{eq:quantum_MC_discrete}, and we then correct our result by inserting the result of the prediction in the right-hand sides of the following equations of motion:
\begin{equation} \label{eq:quantum_MC_discrete2}
\left\{
    \begin{split}
    \Pi^{(R)}_{n+1,m} &= \Pi^{(R)}_{n,m} - \Delta t \frac{z_n R_{n,m} + z_{n+1} R_{n+1,m} }2 + \delta_{n,m} + \Delta t  \frac{F^{(1)}_{n,m}+F^{(1)}_{n+1,m}}2 
    \; , 
    \\
    \Pi^{(C)}_{n+1,m} &= \Pi^{(C)}_{n,m}- \Delta t \frac{z_n C_{n,m} + z_{n+1} C_{n+1,m} }2 + \delta_{n,m}+ \Delta t \frac{F^{(2)}_{n,m}+F^{(2)}_{n+1,m}}2
    \; , 
    \\
    R_{n+1,m} &= R_{n,m} + \Delta t\frac{\Pi^{(R)}_{n,m} + \Pi^{(R)}_{n+1,m}}{2m} 
    \; ,
    \\ 
    C_{n+1,m} &= C_{n,m} + \Delta t\frac{\Pi^{(C)}_{n,m} + \Pi^{(C)}_{n+1,m}}{2m}
    \; .
    \end{split}
\right.
\end{equation}
For each $n$-th step, we take a loop over the predictor-corrector scheme $\mathcal{N}_L$ times. In this way, the error we make is of order $n_{steps}\Delta t^{(1+\mathcal{N}_L)}$. For the results presented in the Letter, we have always fixed $\mathcal{N}_L=2$.

At this point, we observe that Eqs.~\eqref{eq:quantum_MC_discrete2} alone still do not determine the Lagrange multiplier $z_n$, which is in principle determined by the non-causal equation (equivalent to Eq.~\eqref{eq:constraint_literature} of the Letter):
\begin{equation}\label{eq:lagrange_multiplier_discrete}
     z_n  = - \frac{\Pi^{(C)}_{n+1,n}-\Pi^{(C)}_{n,n}}{\Delta t}
     + \Delta t F^{(2)}_{n,m} \ .
\end{equation}
As long as the system is coupled to a thermal bath, we solve this issue by making the physical assumption that, due to dissipation, $z_n$ converges at large times to a stationary value, as also observed in Ref.~\cite{cugliandolo1999real}. Due to this asymptotic convergence, we can safely replace the difference $\Pi^{(C)}_{n+1,n}-\Pi^{(C)}_{n,n}$ with the one evaluated at the previous time step, $\Pi^{(C)}_{n,n}-\Pi^{(C)}_{n,n-1}$, in Eq.~\eqref{eq:lagrange_multiplier_discrete}.
However, this substitution is not valid for $t>t_b$, where the dynamics is isolated and periodically driven. In the latter scenario, we proceed as discussed in Section~\ref{subsec:SM_lagrange_mult} and include the discretized version of Eq.~\eqref{eq:multiplier_eq} in the system of Eqs.~\eqref{eq:quantum_MC_discrete}, so that also $z_n$ can be computed using the predictor-corrector algorithm.

\begin{figure}[ht]
    \centering
    \includegraphics[width=\textwidth]{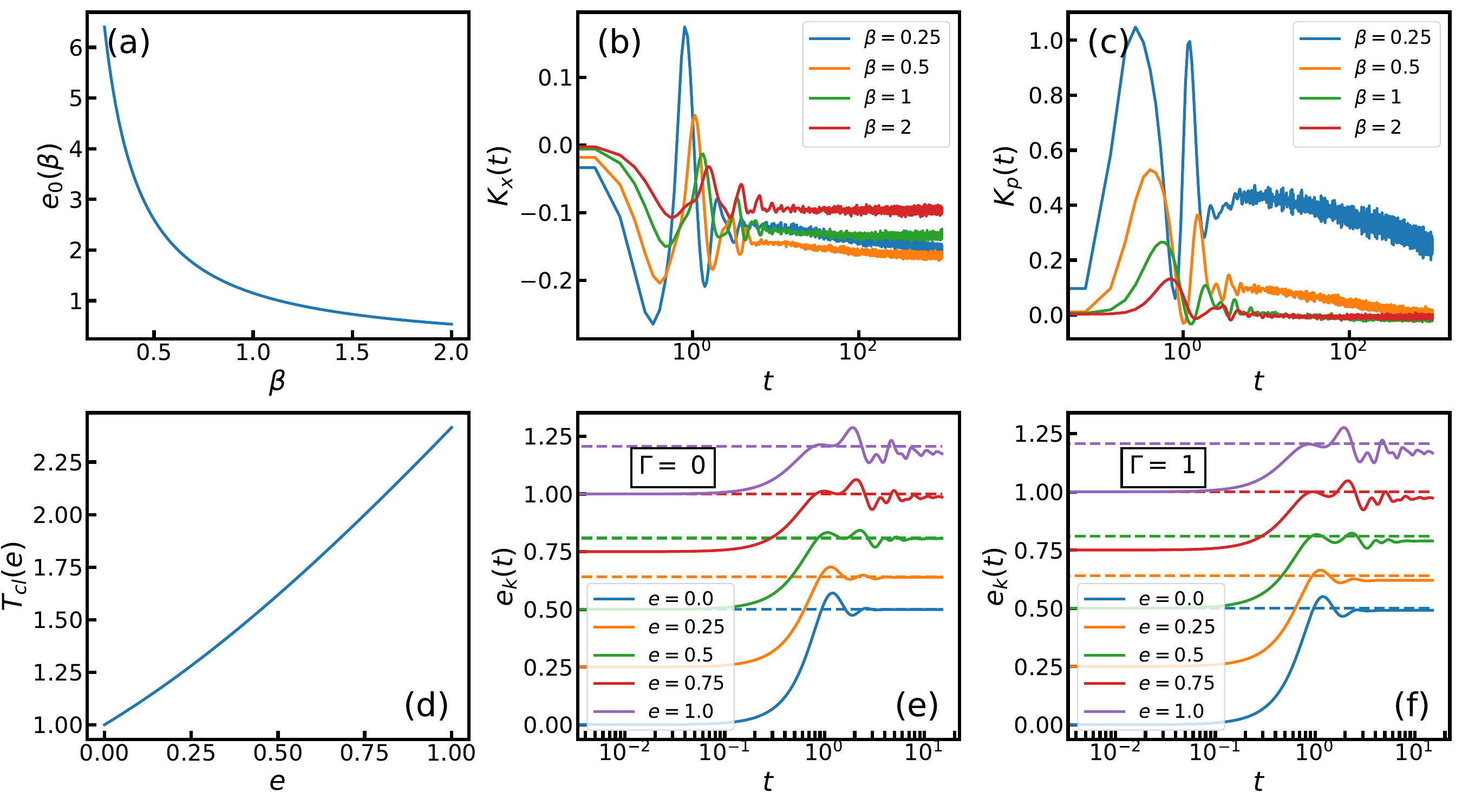}
     \caption{(Color online).
     (a-c) Dynamics of the time-independent 1D lattice $\phi^4$ model. We fix $N=100$ and $\omega_0^2=g=\lambda=1$. (a) Initial energy density as a function of the inverse temperature, from Eq.~\eqref{eq:SM_phi4_quench_initial_energy}. (b-c) Kurtosis of the position and momentum degrees of freedom, from Eq.~\eqref{eq:SM_kurtosis}. Data are averaged over $\mathcal{N}=2000$ configurations.
    (d-f) Dynamics of the time-independent $p$-spin spherical model. (d) Equilibrium dependence of temperature from the energy density $e$, from Eq.~\eqref{eq:SM_PSM_classical_temp}. (e-f) Kinetic energy density from Eq.~\eqref{eq:SM_PSM_kinetic_energy}, for the classical ($\Gamma=0$) and quantum ($\Gamma=1$) cases. Dashed lines corresponds to the corresponding classical equilibrium expectation values from Eq.~\eqref{eq:SM_PSM_equil_kinetic_energy}. We fix $M=J=1$ and $p=3$.}
    \label{fig:phi4_SM_quench}
\end{figure}

\section{Ergodic properties of lattice \texorpdfstring{$\phi^4$}{} model and the \texorpdfstring{$p$}{}-spin spherical model in absence of a drive}\label{sec:SM_ergodicity}

In this section, we discuss the ergodic properties of both the 1D lattice $\phi^4$ model  and of the $p$-spin spherical model, in absence of a periodic drive.\\

We begin by focusing on the $\phi^4$ model, whose dynamics is defined by Eq.~\eqref{eq:SM_generic_Hamilton_eq} at $B_0=0$. We initialize the system in a Gaussian Gibbs distribution, corresponding to the $\phi^4$ Hamiltonian at $\lambda=0$ and given by 
\begin{equation}\label{eq:SM_phi4_Gaussian_state}
    \rho_0(\mathbf{x},\mathbf{p}) = \frac{1}{\mathcal{Z}_0} \exp\Big\{-\frac{\beta}{2} \sum_{k=0}^{N-1
    } (\Tilde{p}_k\Tilde{p}_{-k} + \omega_k^2 \Tilde{x}_k\Tilde{x}_{-k}) \Big\}
\end{equation}
Here, the variables
\begin{equation}
    \tilde{x}_k = \frac{1}{\sqrt{N}}\sum_j e^{i\frac{2 \pi k}{N} j} x_j ~ , \qquad \tilde{p}_k = \frac{1}{\sqrt{N}}\sum_j e^{i\frac{2 \pi k}{N} j} p_j ~ ,
\end{equation}
are the Fourier transform of the phase space variables, $\mathcal{Z}_0$ is a normalization factor and $\omega_k = \sqrt{\omega_0^2 + 2g[1-\cos(2\pi k/N)]}$ are the normal mode frequencies of the chain.
We study the non-equilibrium dynamics evolving from $\rho_0(\mathbf{x},\mathbf{p})$, realized through a quench in the coupling $\lambda$.
It is straightforward to show that the initial energy density of the system is given by
\begin{equation}\label{eq:SM_phi4_quench_initial_energy}
    e_0(\beta) = \frac 1N \braket{H_{\phi^4}
    (\mathbf{x},\mathbf{p})} = \frac{1}{\beta} + \frac{3\lambda}{4\beta^2} \left(\frac 1N\sum_{k=0}^{N-1
    }\frac{1}{\omega_k^2} \right)^2\ .
\end{equation}
The two terms on the right-hand side of Eq.~\eqref{eq:SM_phi4_quench_initial_energy} correspond to the averages of the quadratic and the quartic part of the Hamiltonian, respectively. We plot the function $e_0(\beta)$ in Fig.~\ref{fig:phi4_SM_quench}(a), for a range of $\beta$ roughly corresponding to values of $e_0$ used in the Letter. We integrate the dynamics for some values of $\beta$ from the same range and study the time-evolution of the Kurtosis parameters of the phase space degrees of freedom, defined as:
\begin{equation}\label{eq:SM_kurtosis}
    K_x(t) = \frac{\braket{\sum_i x_i(t)^4} }{3\braket{\sum_i x_i(t)^2} } -1 \ , \qquad 
    K_p(t) = \frac{\braket{\sum_i p_i(t)^4} }{3\braket{\sum_i p_i(t)^2} } -1 \ . 
\end{equation}
Both $K_x(t)$ and $K_p(t)$ vanish for a Gaussian distribution~\cite{Joanes1998kurt}, like the one in Eq.~\eqref{eq:SM_phi4_Gaussian_state}, so that $K_x(0)=K_p(0)=0$. We plot both $K_x(t)$ and $K_p(t)$ in Fig.~\ref{fig:phi4_SM_quench}(b), for several values of $\beta$. We observe that $K_x(t)$ and $K_p(t)$ approach an asymptotic value at late times, although the dynamics moderately slows down at low values of $\beta$. In particular, $K_p(t)$ asymptotically vanishes, while $K_x(t)$ does not. This result is compatible with the approach to a thermal state of the $\phi^4$ model with $\lambda>0$, which is Gaussian in the momentum variables, but not in the position ones.\\

Next, we explore the isolated dynamics of the classical and quantum PSM, defined in Eqs.~\eqref{eq:SM_mc_resp} and~\eqref{eq:SM_mc_corr}. We fix $B(t)=0$ and $\mathcal{J}(t)=1$, to eliminate the periodic drive, and set $t_b=0$ in Eqs.~\eqref{eq:SM_PSM_Sigma} and~\eqref{eq:SM_PSM_D}, to isolate the system from the external bath. We initialize the  the Lagrange multiplier $z(t)$, from Eq.~\eqref{eq:constraint_literature}, to $z(0)=2e$. This choice corresponds to a non-equilibrium initial state, uncorrelated with the disorder and characterized by a vanishing potential energy and kinetic energy density equal to $e$~\cite{Cugliandolo_2017}. As in the classical case, $\Gamma\equiv\sqrt{\hbar^2/M}=0$, the equilibrium energy density $e$ is related to the temperature by ~\cite{Cugliandolo_2017}
\begin{equation}\label{eq:SM_PSM_classical_temp}
   T_{cl}(e) = e+\sqrt{e^2+J^2} \ ,
\end{equation}
we integrate the dynamics for a set of $e$ roughly corresponding to the initial temperatures $T_0$ used in the Letter, as it can be inferred from Fig.~\ref{fig:phi4_SM_quench}(d). In Fig.~\ref{fig:phi4_SM_quench}(e), we plot the evolution of the kinetic energy density,
\begin{equation}\label{eq:SM_PSM_kinetic_energy}
    e_k(t) = \frac{1}{N}\sum_{i=1}^N \braket{\dot{\sigma}_i^2(t)} \ ,
\end{equation}
for the classical case $\Gamma=0$. We observe that $e_k(t)$ quickly saturates to a late time plateau, which is compatible with the equilibrium expectation value (dashed lines)
\begin{equation}\label{eq:SM_PSM_equil_kinetic_energy}
    e_k^{(mc)} = \frac{e+\sqrt{e^2+J^2} }{2}.
\end{equation}
Equation~\eqref{eq:SM_PSM_equil_kinetic_energy} obtained from Eq.~\eqref{eq:SM_PSM_classical_temp} and equipartition theorem. This result proves the fast ergodicity of the system, for the range of temperatures considered in the Letter. A similar result is obtained in the quantum case $\Gamma=1$, as shown in Fig.~\ref{fig:phi4_SM_quench}(f), where the kinetic energy density relaxes roughly to the same plateaus. The similarity between the classical and quantum case suggest that quantum effects are negligible for the parameter we considered.

\section{Floquet dynamics with different driving terms}\label{sec:SM_different_drivings}

In this section, we investigate the periodically driven dynamics of the classical lattice $\phi^4$ model and the $p$-spin spherical model (PSM). We demonstrate that the energy density profiles highlighted in the Letter remain qualitatively unchanged even when choosing a different smooth periodic drive. {Then, we discuss the prethermalizzation in the lattice $\phi^4$ model, under a piecewise constant perodic drive.}

\subsection{Smooth driving force}

\begin{figure}[t]
    \centering
    \includegraphics[width=\textwidth]{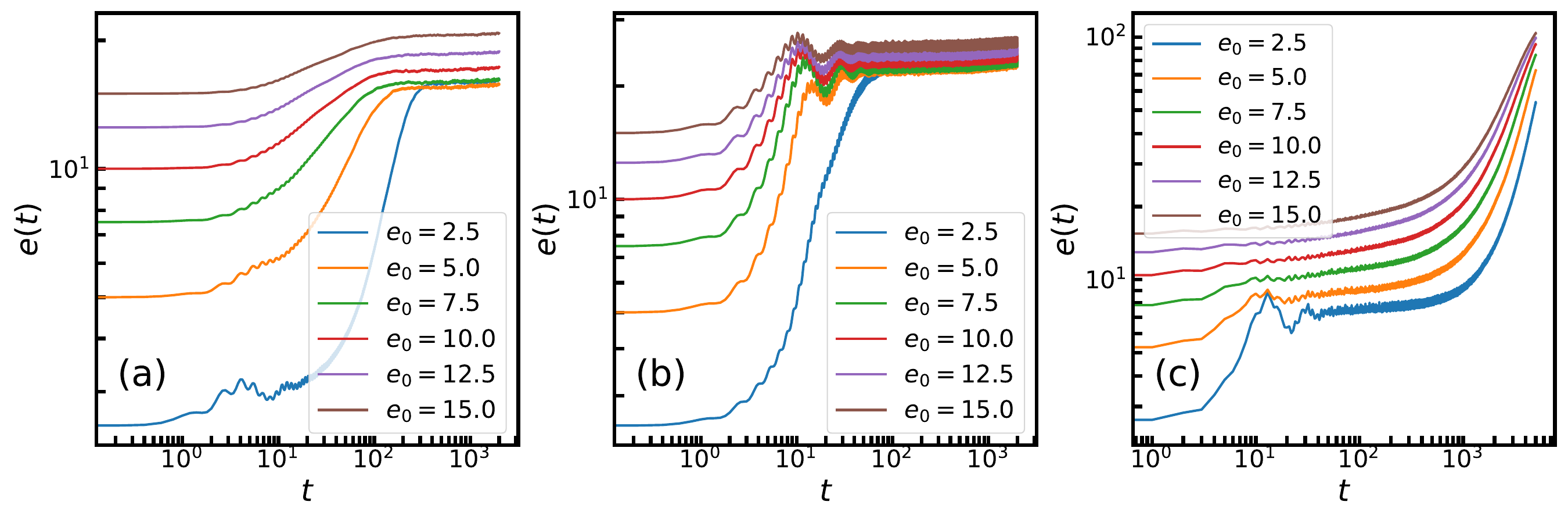}
    \caption{Average energy density  $e(t)=\braket{H_{\phi^4}(\textbf{x}(t),\textbf{p}(t))}_{0}/N$, for the one dimensional lattice $\phi^4$ model, evolving from Eqs.~\eqref{eq:SM_generic_Hamilton_eq_gen}. We fix  $N=100$, $\Omega=2.3$ and $a=\omega_0^2=g=1$. The average displayed here is performed over $\mathcal{N}$ configurations, sampled from a microcanonical manifold at various energy density $e_0$, for the values of $e_0$ listed in the legends. We fix $\mathcal{N}=1000$ and $\lambda=1$ in panels (a) and (b), while $\mathcal{N}=500$ and $\lambda=0.1$ in panel (c). Each panel corresponds to a different form of drive $F_i(\mathbf{x})$. (a) $F_i(\mathbf{x})=\cos(x_i)$. (b) $F_i(\mathbf{x})=x_i^2$. (c) $F_i(\mathbf{x})=x_i$.}
    \label{fig:phi4_SM_other_drives}
\end{figure}

We begin with the lattice $\phi^4$ model, the Hamilton equations of motion of which are given by
\begin{equation}\label{eq:SM_generic_Hamilton_eq_gen}
\ddot{x}_i + \omega_0^2 x_i + \lambda x_i^3 = a \sin( \Omega t) F_i(\mathbf{x}) + g(x_{i+1} -2 x_i + x_{i-1}) \ ,
\end{equation}
where $i=1\ldots N$. If we set $F_i(\mathbf{x})=1$, we recover the same Floquet dynamics examined in the Letter. Instead, here we focus on driving terms corresponding to one of the following possibilities: $F_i(\mathbf{x})=x_i^2$ or $F_i(\mathbf{x})=\cos (x_i)$. We numerically integrate Eqs.~\eqref{eq:SM_generic_Hamilton_eq_gen} over an ensemble of $\mathcal{N}$ initial configurations $\{\mathbf{x}(0),\dot{\mathbf{x}}(0)\}$, randomly sampled on the manifold $H_{\phi^4}\big(\mathbf{x}(0),\dot{\mathbf{x}}(0)\big) = N e_0$, for several values of $e_0$. 
The resulting average energy density, $e(t)=\braket{H_{\phi^4}(\textbf{x}(t),\dot{\textbf{x}}(t))}_0/N$, is depicted in Fig.~\ref{fig:phi4_SM_other_drives}(a) and (b). For both driving terms, we consistently observe that $e(t)$ saturates at long times to a finite value $e_\infty$, underscoring the robustness of the results presented in the Letter across different driving forms.
In Fig.~\ref{fig:phi4_SM_other_drives}(c)  we examine a third, more subtle case involving a periodic modulation of $\omega_0^2$, achieved by setting $F_i(\mathbf{x})=x_i$. The key observation here is that, for a low initial energy $e_0$, the average energy density $e(t)$ exhibits slow but persistent growth within our simulation time scales. This profile, also observed in the closely related $O(N)$ model under the same driving protocol~\cite{Knapp2017prethermal_keldysh}, does not provide a definitive indication of either unbounded growth or saturation to a finite value $e_\infty$. However, we note that the heating diminishes with an increase in the initial energy and eventually disappears for a sufficiently high $e_0$. This observation may indicate the existence of an energy threshold at which heating is expected to cease, leading to the saturation of $e(t)$ to a finite value even when evolving from a low initial energy. To validate this conclusion, further investigation is required.

\begin{figure}[t]
    \centering
    \includegraphics[width=\textwidth]{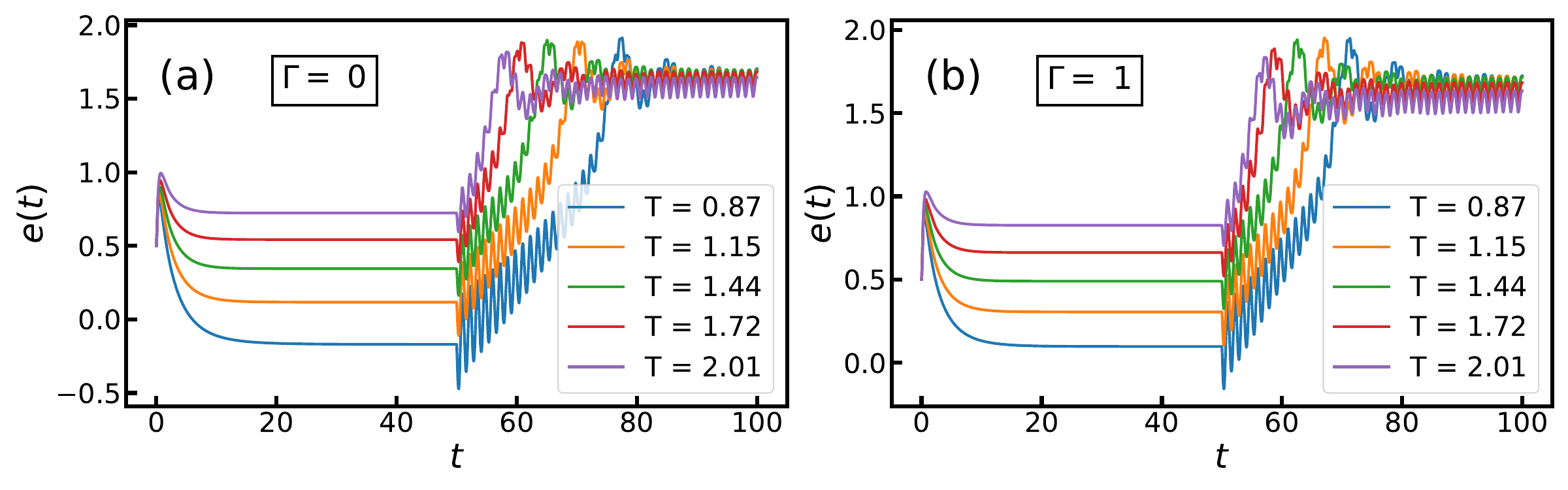}
    \caption{Average energy density  $e(t)$ for the $p$-spin spherical model, defined in Eq.~\eqref{eq:SM_mc_energy} and evolved using Eq.~\eqref{eq:SM_mc_resp} and Eq.~\eqref{eq:SM_mc_corr}. We drive the system by setting $B(t)=0$ and $\mathcal{J}(t)=1+\Delta \sin(\Omega t)$, for $\Delta=0.5$ and $\Omega=5$. We also fix $p=3$, $M=J=1$, $t_b = 50$. Each panel corresponds to a different value of the dimensionless parameter $\Gamma=\sqrt{\hbar^2/M}$
    which quantifies the strength of quantum fluctuations. 
    The different curves in each panel correspond to different initial conditions characterized by their temperature $T_0$ specified in the keys.}
    \label{fig:PSM_SM_other_drives}
\end{figure}

We conduct a similar analysis for the PSM, whose dynamics is governed by the equations of motion~\eqref{eq:SM_mc_resp} and~\eqref{eq:SM_mc_corr}. In the Letter, we investigated Floquet dynamics driven by a homogeneous field, achieved by setting $B(t) = B_0\sin(\Omega t)$ and $\mathcal{J}(t)=1$ in the Hamiltonian in Eq.~\eqref{eq:SM_driven_hamiltonian_SG}. Here, our focus shifts to a modulation in the interaction term, achieved by setting $B(t)=0$ and $\mathcal{J}(t)=1+\Delta \sin(\Omega t)$. As illustrated in Fig.~\ref{fig:PSM_SM_other_drives}, the corresponding energy density, Eq.~\eqref{eq:SM_mc_energy}, saturates at long times to a finite value, consistently with the results presented in the Letter for a different kind of drive.

\subsection{Piecewise constant driving force}

{In this section, we investigate the dynamics of the lattice $\phi^4$ model, under a piecewise constant periodic drive. The corresponding Hamilton equations of motion of are given by:
\begin{equation}\label{eq:SM_Hamilton_eq_step}
\ddot{x}_i + \omega_0^2 x_i + \lambda x_i^3 = a     
~\theta(\sin( \Omega t)) F_i(\mathbf{x}) + g(x_{i+1} -2 x_i + x_{i-1}) \ ,
\end{equation}
where $i=1\ldots N$ and $\theta(x)$ is the Heaviside step function. We focus on the case where either $F_i(\mathbf{x)}=1$ or $F_i(\mathbf{x)}=x_i^2$ and compute the average energy density $e(t)$ as we did in the previous section. The results are shown in Fig.~\ref{fig:phi4_SM_step_drive}. 
Comparing these results with those obtained both in the Letter and in the previous section, we observe that the prethermal plateau at $e_\infty$ is shorter for a discontinuous drive than for a smooth one. Furthermore, the range of driving frequencies where a prethermal plateau is observed is also narrower.}

\begin{figure}[hb]
    \centering
    \includegraphics[width=\textwidth]{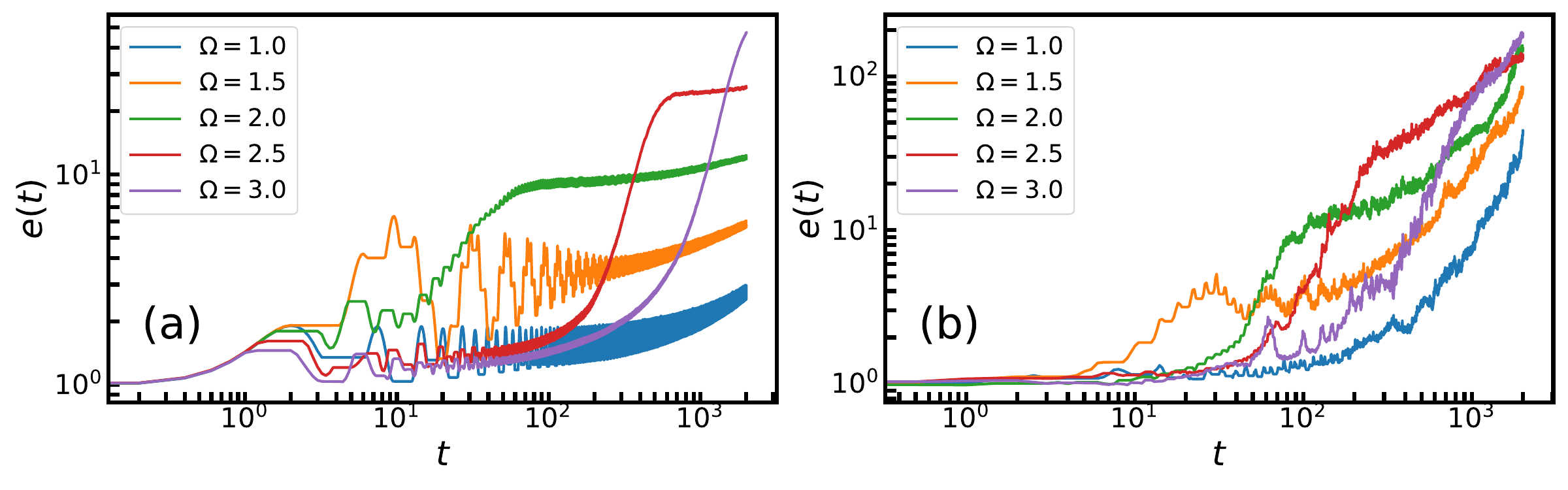}
    \caption{{Average energy density  $e(t)=\braket{H_{\phi^4}(\textbf{x}(t),\textbf{p}(t))}_{0}/N$, for the one dimensional lattice $\phi^4$ model under a step drive, derived from  Eqs.~\eqref{eq:SM_Hamilton_eq_step}. We fix $e_0=1$ and $\Omega$ as specified in the legends. The dynamics is averaged over $\mathcal{N}=1000$ initial configurations. The drive is characterized by $F_i(\mathbf{x})=1$ in (a) and $F_i(\mathbf{x})=x_i^2$ in (b). All other parameters are fixed as in Fig.~\ref{fig:phi4_SM_other_drives}.}}
    \label{fig:phi4_SM_step_drive}
\end{figure}

\end{widetext}

\end{document}